\documentclass[aps,prb,twocolumn,floatfix,showpacs,citeautoscript,longbibliography,hyperlinks]{revtex4-1}

\usepackage{graphicx}
\usepackage{dcolumn}
\usepackage{bm}
\usepackage{bbold}
\usepackage{mathtools}
\usepackage{color}
\usepackage{transparent}
\usepackage[colorlinks=true,citecolor=blue]{hyperref}

\newcommand{\revision}[1]{{{#1}}}

\begin{document}

\title{\revision{Optimization of quantized charge pumping using full counting statistics}}
\author{Elina Potanina}
\author{Kay Brandner}
\author{Christian Flindt}
\affiliation{
 Department of Applied Physics, Aalto University, 00076 Aalto, Finland}

\date{\today}

\begin{abstract}
We \revision{optimize the operation of single-electron charge pumps using full counting statistics techniques}. \revision{To this end, we evaluate the statistics of pumped charge on a wide range of driving frequencies using Floquet theory, focusing here on the current and the noise}. For charge pumps controlled by one or two gate voltages, we demonstrate that our theoretical framework \revision{may lead to enhanced device performance}. Specifically, by optimizing the driving parameters, we predict a significant increase in the frequencies for which a quantized current can be produced. For adiabatic two-parameter pumps, \revision{we exploit that the pumped charge and the noise can be expressed as surface integrals over Berry curvatures in parameter space.} Our findings are important for the efforts to realize high-frequency charge pumping, and our predictions may be verified using current technology.
\end{abstract}

\maketitle

\section{Introduction} 
\begin{figure}
\centering
\includegraphics[width=0.46\textwidth]{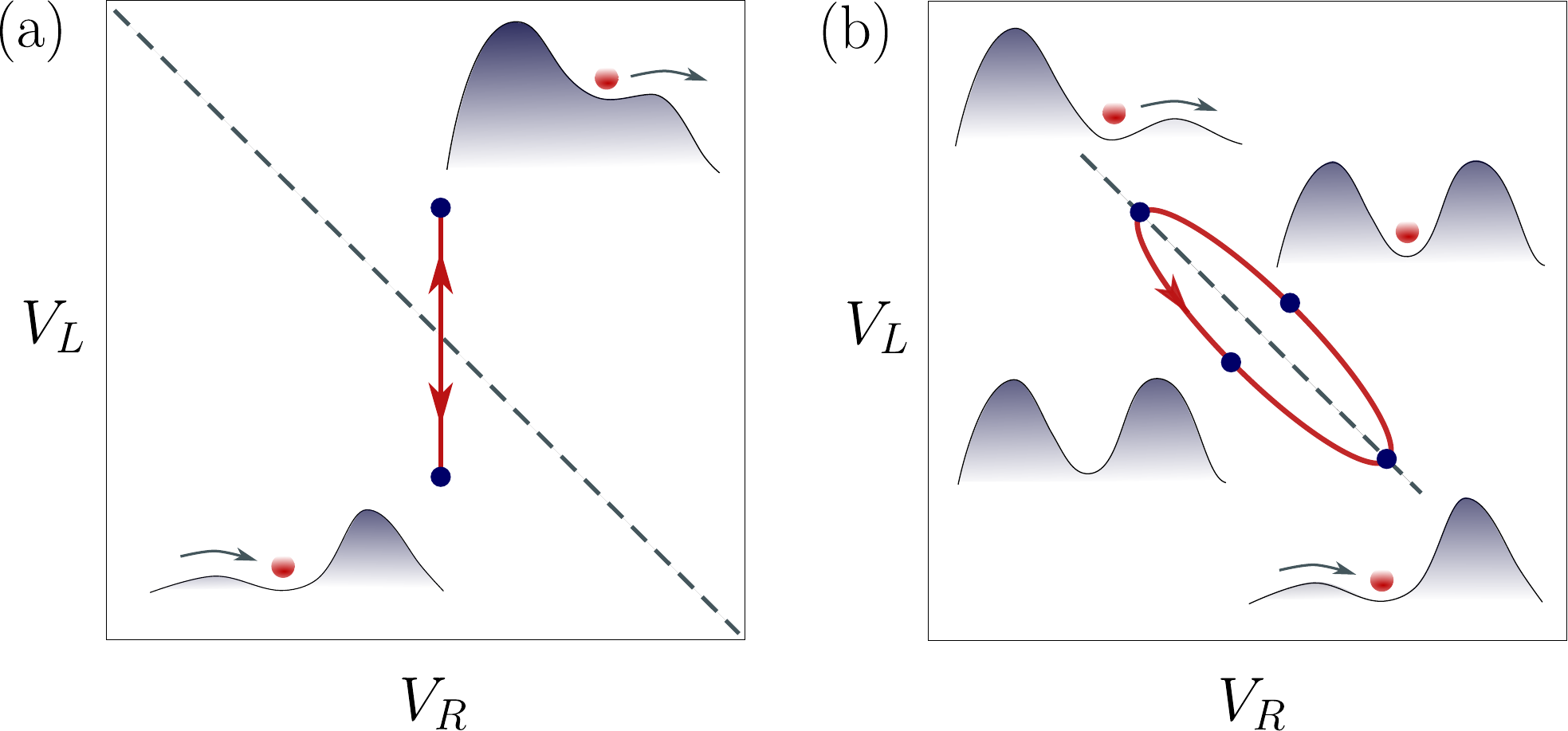}
\caption{Single-electron pumping. (a) Non-adiabatic charge pumping can be achieved by modulating a single gate voltage periodically in time as indicated by the red line. \revision{In this case, mainly the left barrier of the gate-defined potential is periodically modulated as illustrated by the insets.} The dashed line separates the stable charge configurations of the island (0 or 1 electrons). (b) Adiabatic pumping can be achieved by slowly modulating both gate voltages periodically in time as indicated by the positively-oriented contour in red. \revision{The insets illustrate how both barriers are periodically modulated.}}
\label{fig:Pump}
\end{figure}

Single-electron pumps are important for a wide range of quantum technologies, and they have been proposed as precise current sources for metrological purposes \cite{Odintsov1991,Giblin2012,Pekola2013}. The central goal is to transfer single electrons between two leads via a nano-scale island as accurately and as fast as possible. The gate voltages of the island are modulated periodically in time with the aim to generate a current given by the electron charge times the frequency of the drive, Fig.~\ref{fig:Pump}. Single-electron pumping has been demonstrated in several experimental architectures, and both the accuracy and the driving speed have been significantly increased during recent years \cite{Odintsov1991,Giblin2012,Pekola2013,Kouwenhoven1991,Pothier1992,Keller1996,Ono2003,Robinson2005,Pekola2007,Fujiwara2008,Jehl2013,Yamahata2014,Rossi2014,Connolly2013,Blumenthal2007,SlavaBernd2008,Giblin2010,Kataoka2011,Fricke2014,Ubbelohde2014,KaestnerKashcheyevs2015,Stein2015,Yamahata2016,Ahn2017,Zhao2017,BrunPicard2016}.

To achieve reliable loading and unloading of single electrons, it is generally favorable to operate the pumps at low frequencies \cite{Moskalets2002,Moskalets2002b}. This regime can be elegantly described using adiabatic theories~\cite{Brouwer1998,Aleiner98,Shutenko2000,Avron2000,MakhlinMirlin2001, EntinWohlman2002,Splettstoesser2005}. However, to produce an appreciable current, the driving should be fast, while maintaining faultless single-electron control. Moreover, pumps operating with a single modulated gate voltage only deliver a quantized current well beyond the adiabatic regime \cite{Blumenthal2007,SlavaBernd2008,Giblin2010,Kataoka2011,Fricke2014,Ubbelohde2014,KaestnerKashcheyevs2015,Stein2015,Yamahata2016,Ahn2017,Zhao2017}. Various techniques have been developed to improve the accuracy of such non-adiabatic pumps at the quantized-current plateau~\cite{Kashcheyevs2010, Kashcheyevs2012}. On the other hand, efficient tools to optimize the driving frequency are still lacking, as it is challenging to develop theories that extend beyond the adiabatic approximation. Instead, non-adiabatic pumps have mainly been investigated using numerical approaches~\cite{SlavaBernd2008,Ohkubo2010,Croy2012,Croy2016}.

In this work, we \revision{employ full counting statistics techniques to optimize the operation of single-electron charge pumps. We use Floquet theory to evaluate the current and the fluctuations of the pumped charge} order-by-order in either the frequency or the period of the drive and thereby develop a systematic understanding of charge pumps beyond the adiabatic approximation. For single-parameter pumps, we optimize the driving frequency by minimizing the noise over the pumped charge (the Fano factor) at high frequencies. For adiabatic  pumps, the full counting statistics can be expressed \revision{as a surface integral over a Berry curvature in parameter space~\cite{Sinitsyn2007Berry,Sinitsyn2007,Hanggi2010,Goswami2016}, which we use to optimize the driving protocol.} Moreover, from the high-frequency expansion we can estimate the breakdown frequency for which a quantized current can no longer be generated. \revision{Although, we focus here on the} average and noise of the pumped charge, our theoretical framework is versatile, and it can readily be adapted to other quantities \revision{such as the higher cumulants} or even the large-deviation statistics of the current~\cite{TOUCHETTE2009}.

\section{Quantized charge pumping} The Floquet theory that we develop below is applicable to a large class of open quantum systems that exchange particles (or heat) with external reservoirs and whose dynamics can be described by a Markovian (generalized) master equation. To be specific, we here consider periodically-driven single-electron pumps that ideally transfer one electron from a source electrode to a collector in every single operation cycle. A charge pump consist\revision{s} of a nano-scale island whose dynamics is governed by the master equation
\begin{equation}
      \frac{d}{dt}\vert  P(t) \rangle = \mathbf{L}(t)\vert  P(t) \rangle,
\label{eq:RateEq}
\end{equation}
where the vector $\vert  P(t) \rangle=[p_0(t),p_1(t),p_2(t),\ldots]^{T}$ contains the probabilities for the island to be occupied by $0,1,2,...$ electrons. The rate matrix $\mathbf{L}(t)=\mathbf{L}(t+\mathcal{T})$ describes the transitions between different charge states of the island, and $\mathcal{T}$ is the period of the external drive. At all times, the product of the tunneling amplitudes to the source and the collector is kept so small that co-tunneling processes can safely be ignored and we may consider sequential single-electron tunneling only.

To investigate the pumped current, we resolve the probability vector $\vert  P(t) \rangle=\sum_n \vert  P(n,t) \rangle$ with respect to the number of electrons $n$  that have been transferred during the time-span $[0,t]$ \cite{Plenio1998}. The charge transfer statistics can then be expressed as $P(n,t)=\langle 1 \vert P(n,t) \rangle$ with all entries of the vector $\langle 1 \vert$ being~$1$. We also write the rate matrix as $\mathbf{L}(t)=\mathbf{L}_0(t)+\mathbf{J}_+(t)+\mathbf{J}_{-}(t)$ with $\mathbf{J}_{\pm}(t)$ describing charge transfers to and from the collector \cite{Benito2016}. The equations of motion, $\frac{d}{dt}\vert  P(n,t) \rangle =\mathbf{L}_0(t)\vert  P(n,t) \rangle +\mathbf{J}_+(t)\vert  P(n-1,t) \rangle +\mathbf{J}_{-}(t)\vert  P(n+1,t) \rangle$, are decoupled by introducing the counting field $\chi$ via the definition $\vert  P(\chi,t) \rangle\equiv\sum_n\vert  P(n,t) \rangle e^{in\chi}$. We then arrive at a modified master equation for $\vert  P(\chi,t)\rangle$
\begin{equation}
      \frac{d}{dt}\vert  P(\chi,t) \rangle = \mathbf{L}(\chi,t)\vert  P(\chi,t) \rangle
\label{eq:RateEq_chi}
\end{equation}
with $\mathbf{L}(\chi,t)=\mathbf{L}(t)+(e^{i\chi}-1)\mathbf{J}_+(t)+(e^{-i\chi}-1)\mathbf{J}_{-}(t)$. Formally, the solution $\vert  P(\chi,t) \rangle= \mathbf{U}(\chi,t)\vert  P(\chi,0) \rangle $ is given by the time-ordered exponential  $\mathbf{U}(\chi,t)=  \hat{T} \{e^{\int_{0}^{t} dt' \mathbf{L}(\chi,t')}\}$ \cite{Pistolesi2004,Potanina2017}.  The moments of the pumped charge~$n$ then follow as $\langle n^m\rangle (t)=\partial_{i\chi}^m \mathcal{M}(\chi,t)|_{\chi=0}$, where $\mathcal{M}(\chi,t)\equiv\sum_n P(n,t)e^{in\chi}=\langle 1 \vert\mathbf{U}(\chi,t)\vert  P(\chi,0) \rangle$ is the moment generation function. Similarly, the cumulant generating function $\mathcal{S}(\chi,t)\equiv\ln \mathcal{M}(\chi,t)$ delivers the cumulants as $\langle\!\langle n^m\rangle\!\rangle (t)=\partial_{i\chi}^m \mathcal{S}(\chi,t)|_{\chi=0}$. Below, we \revision{focus on} the first two cumulants, namely the mean $\langle\!\langle n\rangle\!\rangle =\langle n\rangle$ and the variance $\langle\!\langle n^2\rangle\!\rangle =\langle n^2\rangle-\langle n\rangle^2$, \revision{although higher cumulants can easily be obtained with little added effort.}

\section{Floquet theory} We now make use of the periodicity of the drive. Building on the Floquet theorem \cite{BukovPolkovnikov2015}, the time-evolution operator can be expressed as $\mathbf{U}(\chi,t)=\sum_i e^{\lambda_i(\chi)t}  \vert p_i(\chi,t)\rangle\!\langle p_i(\chi,0)\vert$, where $\vert  p_i(\chi,t) \rangle=\vert  p_i(\chi,t+\mathcal{T}) \rangle$ solves the Floquet eigenvalue problem \footnote{We note that the left and right eigenvectors, $\protect\langle p_i(\chi,t)\protect\vert$ and $\protect\vert p_i(\chi,t)\protect\rangle$, are not related by simple Hermitian conjugation, since the rate matrix $\protect\mathbf{L}(\chi,t)$ is not Hermitian.}
\begin{equation}\label{eq:EigenEq}
    \left[\mathbf{L}(\chi,t) -  \frac{d}{dt} \right] \vert p_i(\chi,t)\rangle = \lambda_i(\chi) \vert p_i(\chi,t)\rangle.
\end{equation}
We then obtain $\mathcal{S}(\chi,t)=\ln{\sum_i e^{\lambda_i(\chi)t} \langle 1 \vert p_i(\chi,t)\rangle}$ and immediately see that the charge transfer statistics after many periods $\mathcal{N}\gg 1$ is fully encoded in the Floquet eigenvalue $\phi(\chi)\equiv \max\limits_i \left[ \lambda_i (\chi) \right]$ with the largest real-part
\begin{equation}
 \mathcal{S}(\chi,\mathcal{N} \mathcal{T}) \simeq \mathcal{N} \mathcal{T}\phi(\chi).
 \label{eq:CGFmax}
\end{equation}
Generally, however, it is a daunting task to determine $\phi(\chi)$ and its dependence on the counting field. Nevertheless, as we go on to show, the eigenvalue can be found perturbatively in the frequency or the period of the drive.

\begin{figure*}
\centering
\includegraphics[width=0.95\textwidth]{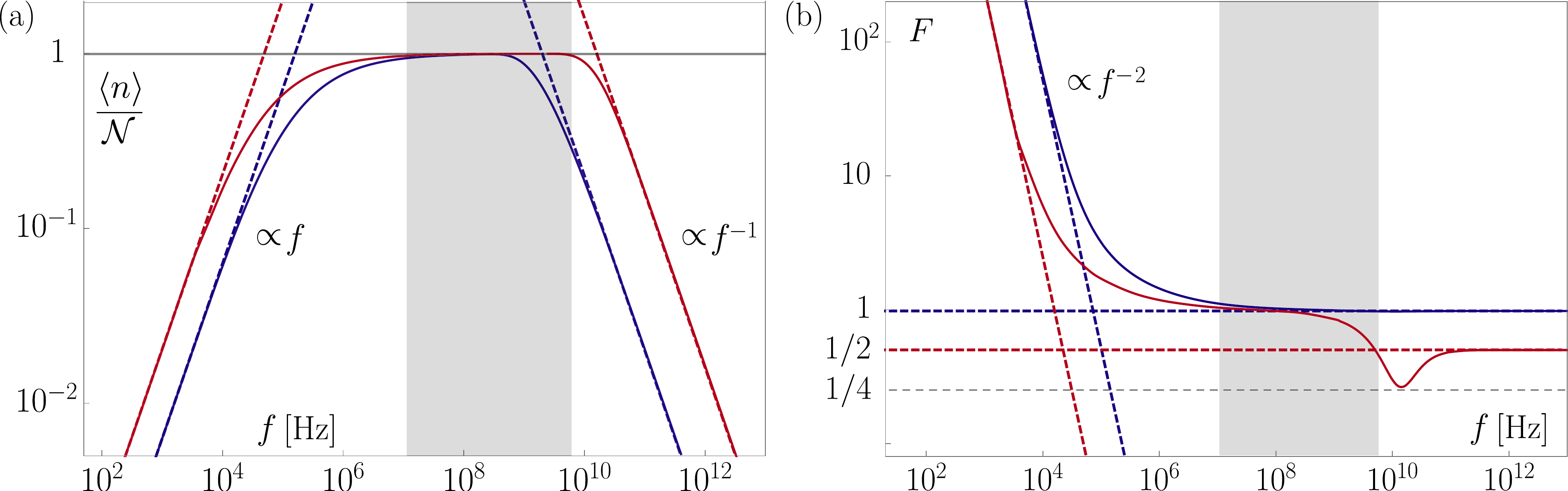}
\caption{Single-parameter pump. (a) Average pumped charge per period as a function of the driving frequency $f$. The solid lines are numerical results, while the dashed lines are the low and high-frequency expansions. The red line is obtained with driving parameters that minimize the Fano factor at high frequencies. The shaded area indicates the quantized-current plateau for the optimized driving parameters. The system parameters are $C=10$~aF,  $G=1.0\times10^{-4}$~$\Omega^{-1}$, $T=0.2$~K, and $\Gamma=G/4C = 2.5\times 10^{12}$~s$^{-1}$. The driving parameters are $V_L^0 = 4$~mV (blue), $24$~mV (red), $V_R^0=0.8$~mV, $V_s=5$~$\mu$V, $\mathcal{N}_g=0.2$.  (b) The Fano factor $F=\langle\!\langle n^2\rangle\!\rangle/\langle n\rangle$ of the pumped charge as a function of the driving frequency.}
\label{fig:oneparameter_pump}
\end{figure*}

\section{Adiabatic expansion} We first evaluate the Floquet eigenvalue $\phi(\chi)$ and the corresponding eigenvector, denoted as $\vert p(\chi,t)\rangle$, perturbatively in the driving frequency. In the adiabatic expansion, we treat the time-derivative $-\frac{d}{dt}$ in Eq.~(\ref{eq:EigenEq}) as the perturbation \cite{Cavina2017}. Our adiabatic expansion can be formulated in terms of the instantaneous eigenvalue of  $\mathbf{L}(\chi,t)$ with the largest real-part $\lambda^{(0)}(\chi,t)$ and the corresponding eigenvectors $\langle p^{(0)}(\chi,t)\vert$ and $\vert p^{(0)}(\chi,t)\rangle$.
To begin with, we find from Eq.~({\ref{eq:EigenEq}})
\begin{equation}\label{eq:EigenValue}
      \phi(\chi) = \phi^{(0)}(\chi)-\int_0^{\mathcal{T}} \frac{dt}{\mathcal{T}} \langle p^{(0)}(\chi,t) \vert \frac{d}{dt}\vert p(\chi,t)\rangle,
\end{equation}
where $\phi^{(0)} (\chi) =\int_0^{\mathcal{T}} \frac{dt}{\mathcal{T}} \lambda^{(0)}(\chi,t)$ is the average of the instantaneous eigenvalue. Without a voltage bias, the contribution to  the mean current from $\phi^{(0)}(\chi)$ vanishes and the noise can be related to the conductance according to the fluctuation-dissipation theorem \cite{Blanter2000}. To proceed to higher orders, we expand the eigenvalue and eigenvector in the perturbation as $\phi(\chi)=\sum_{k=0}^{\infty}\phi^{(k)}(\chi)$ and $\vert p(\chi,t)\rangle=\sum_{k=0}^{\infty}\vert p^{(k)}(\chi,t)\rangle$ and collect terms of the same order in Eq.~(\ref{eq:EigenValue}). To first order, we find $ \phi^{(1)}(\chi)= - \int_0^{\mathcal{T}} \frac{dt}{\mathcal{T}}  \langle p^{(0)}(\chi,t) \vert \frac{d}{dt} \vert  p^{(0)}(\chi,t) \rangle$ as previously established within a different framework \cite{Sinitsyn2007Berry,Sinitsyn2007,Hanggi2010,Goswami2016}. For a device controlled by a single parameter, this term vanishes as we discuss below. To second order, we find
\begin{equation}
    \phi^{(2)} (\chi) = -\int_0^{\mathcal{T}} \frac{dt}{\mathcal{T}}  \langle p^{(0)}(\chi,t) \vert \frac{d}{dt}\mathbf{R}(\chi,t)\frac{d}{dt} \vert  p^{(0)}(\chi,t) \rangle
    \label{eq:na_term}
\end{equation}
having used $\vert  p^{(1)}(\chi,t) \rangle=\mathbf{R}(\chi,t)\frac{d}{dt} \vert  p^{(0)}(\chi,t)\rangle$ as in standard perturbation theory, where  $\mathbf{R}(\chi,t)$ is the pseudo-inverse of $\mathbf{L}(\chi,t)-\lambda^{(0)}(\chi,t)$ \cite{Flindt2010}. Equation~(\ref{eq:na_term}) is important as it allows us to evaluate the charge transfer statistics for single-parameter pumps to first non-trivial order in the driving frequency. Before demonstrating its usefulness with specific applications, we discuss our high-frequency expansion of the cumulant generating function.

\section{High-frequency expansion} The high-frequency expansion proceeds differently. Here, we write the time-evolution operator as $\mathbf{U}(\chi,\mathcal{N}\mathcal{T})=[\mathbf{U}(\chi,\mathcal{T})]^\mathcal{N}:=e^{\mathcal{N}\mathcal{T}\mathbf{L}_F(\chi)}$ and identify the Floquet eigenvalue $\phi(\chi)$ as the eigenvalue of  $\mathbf{L}_F(\chi)$ with the largest real-part. Using a Floquet-Magnus expansion $\mathbf{L}_F(\chi)=\sum_{k=0}^{\infty}\mathbb{L}^{(k)}(\chi)$, we can then evaluate $\phi(\chi)$ perturbatively in the period. The first two terms read $\mathbb{L}^{(0)}(\chi) = \int_0^{\mathcal{T}} \frac{dt}{\mathcal{T}} \mathbf{L}(\chi,t)$
and $\mathbb{L}^{(1)}(\chi) = \int_0^{\mathcal{T}} \frac{dt}{2} \int_0^{t}  \frac{dt'}{\mathcal{T}} [ \mathbf{L}(\chi,t), \mathbf{L}(\chi,t')]$ \cite{Blanes2009,BukovPolkovnikov2015,KUWAHARA201696}. In the high-frequency expansion $\phi(\chi)=\sum_{k=0}^\infty \varphi^{(k)}(\chi)$, the first term $\varphi^{(0)}(\chi)$ is given by the eigenvalue of $\mathbb{L}^{(0)}(\chi)$ with the largest real-part. Denoting the corresponding eigenvectors by  $\langle \mathbb{p}^{(0)}(\chi)\vert$ and $\vert \mathbb{p}^{(0)}(\chi)\rangle$, the next term becomes
    $\varphi^{(1)} (\chi) =   \langle \mathbb{p}^{(0)}(\chi)\vert \mathbb{L}^{(1)}(\chi)\vert \mathbb{p}^{(0)}(\chi)\rangle.$
Thus, with the eigenvectors of $\mathbb{L}^{(0)}(\chi)$ at hand, we can evaluate the charge transfer statistics perturbatively in the period.

\section{Single-electron pump} We can now analyze a charge pump which is similar to those from recent experiments \cite{Giblin2012,Ono2003,Fujiwara2008,Jehl2013,Yamahata2014,Rossi2014,Blumenthal2007,Giblin2010,Kataoka2011,Fricke2014,Ubbelohde2014,Stein2015,Yamahata2016,Ahn2017,Zhao2017}. The pump consists of a metallic island operated in the Coulomb-blockade regime, where the island is either empty or occupied by one electron. The rate matrix then takes the simple form
\begin{equation}
    \mathbf{L} (\chi, t) = \left( \begin{array}{ll}
        -\Gamma^{+}_L (t) - \Gamma^{+}_R (t)  & \Gamma^{-}_L (t) + \Gamma^{-}_R (t) e^{i \chi}  \\\\
         \Gamma^{+}_L (t) + \Gamma^{+}_R (t) e^{-i \chi}  & -\Gamma^{-}_L (t) - \Gamma^{-}_R (t)
    \end{array}
    \right),
    \nonumber
\end{equation}
where $\Gamma_\alpha^{\pm}(t)= \frac{G_{\alpha}(t)}{q^2} \frac{\pm \Delta E(t)}{\exp[\pm \beta \Delta E(t)]-1}$ is the rate at which tunneling occurs between the island and the leads, changing the occupation by $\pm 1$ electron with charge $-q$.  No voltage bias is applied, and $\beta=1/k_BT$ is the inverse temperature. The change of the electrostatic energy due to the addition of an electron reads $\Delta E(t) = -E_{\mathrm{c}}[\mathcal{N}_g+2\{C_{L}V_L(t)+C_{R}V_R(t)\}/q]$, where $C_{\alpha}$ are the gate capacitances, $E_{\mathrm{c}}=q^2/2(C_{L}+C_{R})$ is the charging energy, and the offset $\mathcal{N}_g$ can be controlled with a back-gate \cite{Pekola2013}. The barrier conductances depend exponentially on the gate voltages, $G_{\alpha}(t) = G_\alpha \exp[V_{\alpha}(t)/V_s]$, where $V_s$ is known as  the sub-threshold slope \cite{Jehl2013}.

\begin{figure*}
\centering
\includegraphics[width=0.95\textwidth]{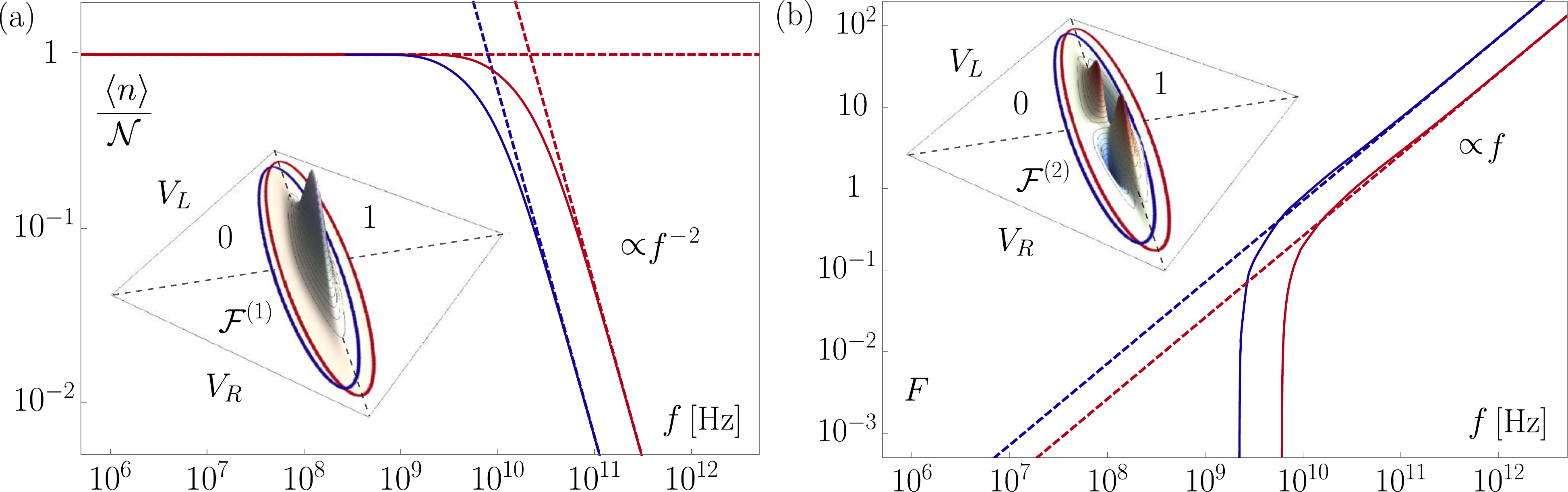}
\caption{Pumped charge (a) and the Fano factor (b) for the two-parameter pump. The driving protocol $\textbf{V}(t) = -V^0[\cos(2 \pi f t)+\alpha, \cos(2 \pi f t+\vartheta)+\alpha]^T$ is shown together with the Berry curvatures $\mathcal{F}^{(j)}$ in the insets,  where the stable charge configuration of the island is also indicated (0 or 1 electrons). The solid lines are numerical results, while the dashed lines are the low and high-frequency approximations. The system parameters are given in Fig.~\ref{fig:oneparameter_pump}. The driving parameters are $V^0=8$ mV, $\vartheta=1.02\pi$,  $\mathcal{N}_g=2$, and $\alpha=1.01$ (blue), $\alpha=\mathcal{N}_g/2=1$ (red).}
\label{fig:CurrentFano}
\end{figure*}

\section{Single-parameter pumping} We first consider a single-parameter pump, where the right gate voltage is kept constant, $V_R(t)=-V_R^0$, while the left one is subject to the harmonic drive $V_L(t)=-V_{L}^{0}\left[\cos(2 \pi f t)+1\right]$. For low frequencies, the average of the pumped charge is obtained from Eq.~(\ref{eq:na_term}). At low temperatures, where the tunneling rates $\Gamma^{-}_L (t)\simeq\Gamma^{+}_R (t)\simeq 0 $ are small, we find
\begin{equation}
 \frac{\langle n \rangle}{\mathcal{N}} \simeq f\int_0^{1} ds \frac{ \left[\Gamma_L^{+}(s)\right]^4 \left(\frac{d}{ds} [\Gamma^{-}_R(s)/\Gamma^{+}_L(s)] \right)^2}{ \left[\Gamma^{+}_L(s) + \Gamma^{-}_R(s)\right]^5},
 \label{nNonGeo}
\end{equation}
having introduced the dimensionless time $s=ft$ to show that the pumped charge is proportional to the driving frequency $f$. We also find that the variance can be expressed as $\langle\!\langle n^2\rangle\!\rangle/\mathcal{N}=2k_BT\int_0^1 ds G(s)/q^2f$ in terms of the instantaneous linear conductance $G(t)$ in accordance with the fluctuation-dissipation theorem. Combined with Eq.~(\ref{nNonGeo}), we see that the Fano factor $F=\langle\!\langle n^2\rangle\!\rangle/\langle n\rangle$ must be proportional to $f^{-2}$ at low frequencies.

For high frequencies, we find the pumped charge from the first term in the Floquet-Magnus expansion,
\begin{equation}
    \frac{\langle n \rangle}{\mathcal{N}} \simeq \frac{\Gamma}{f}  \left[\frac{q e^{V_R^0/V_ s}}{2 C(V_L^0+V_R^0)-q\mathcal{N}_g}+\frac{q \sqrt{2\pi V_L^0/V_s}}{q\mathcal{N}_g-2C V_R^0}\right]^{-1}.
     \label{DynCurr}
\end{equation}
Here, we have taken $C_{L}=C_{R}=C$ and $G_L=G_R=G$ with $\Gamma=G/4C$ being an inverse $RC$-time. The gate voltage must change considerably compared to the sub-threshold slope to open and close the left barrier, while being smaller than the charging energy, so that $2CV_R^0 < q \mathcal{N}_g < 2 C(V_L^0+V_R^0)$. The Fano factor thus becomes
\begin{equation}
    F \simeq \frac{e^{-\frac{2V_R^0}{V_s}}[2C(V_L^0+V_R^0)-q\mathcal{N}_g]^2 + \frac{V_s}{2\pi V_L^0} (q\mathcal{N}_g-2CV_R^0)^2 }{\left(e^{-\frac{V_R^0}{V_s}}[2C(V_L^0+V_R^0)-q\mathcal{N}_g] + \sqrt{\frac{V_s}{2\pi V_L^0}}(q\mathcal{N}_g-2CV_R^0) \right)^2}.
    \label{eq:Fano}
\end{equation}

Figure~\ref{fig:oneparameter_pump} shows numerical results for the pumped charge and the Fano factor together with our approximations. The blue curves illustrate the good agreement between the numerics and our perturbative results. With Eqs.~(\ref{nNonGeo},\ref{DynCurr}) we quantitatively explain the low and high frequency dependence of the pumped charge, which previously has been observed in numerical calculations \cite{SlavaBernd2008}.  Moreover, our results allow us to optimize the driving parameters. By inspecting Eq.~(\ref{eq:Fano}), we see that the Fano factor takes the  minimal value of $1/2$, if $e^{-\frac{V_R^0}{V_s}}[2C(V_L^0+V_R^0)-q\mathcal{N}_g] = \sqrt{\frac{V_s}{2\pi V_L^0}}(q\mathcal{N}_g-2CV_R^0)$. We then obtain an optimal ratio of the noise over the pumped charge, which simplifies to $\langle n \rangle/\mathcal{N} \simeq (\Gamma/2f)\sqrt{V_s/2 \pi V_L^0} (\mathcal{N}_g-2 C V_R^0/q) $. The red lines in Fig.~\ref{fig:oneparameter_pump} show the results of this optimization. Importantly, compared to the generic blue curve, we obtain an order-of-magnitude increase in the frequencies, for which a quantized current can be produced. Interestingly, the Fano factor dips below $1/2$ at the end of the quantized-current plateau and almost reaches $1/4$, signaling a transition to a new transport regime.

\section{Two-parameter pumping} Next, we modulate both voltages periodically in time, $\textbf{V}(t) = [V_L(t),V_R(t)]^T$. In the adiabatic regime, we can then write $\phi^{(1)}(\chi)=\pm f \int\!\!\!\int_\mathcal{S} d V_L d V_R \mathcal{F}(\chi,\textbf{V} )$ by virtue of Stoke\revision{s'} theorem. Here, the sign is given by the orientation of the contour enclosing the surface $\mathcal{S}$  in the parameter space, and $
    \mathcal{F}(\chi,\textbf{V}) = \left[-\partial_{V_L},\partial_{V_R}\right]\cdot \langle p^{(0)}(\chi,\textbf{V} ) \vert \nabla_\textbf{V}\vert  p^{(0)}(\chi,\textbf{V} )  \rangle$
is a classical analog of the Berry curvature in quantum mechanics \cite{Sinitsyn2007Berry,Sinitsyn2007,Hanggi2010,Goswami2016}. Clearly, if only one voltage is varied, the surface area vanishes, and $\phi^{(1)}(\chi)=0$.  For the pumped charge  \cite{Pekola1999,Levinson2001}, we find $\langle n \rangle/\mathcal{N} \simeq \pm \int\!\!\!\int_\mathcal{S} d V_L d V_R \mathcal{F}^{(1)}(\textbf{V} )$ with $\mathcal{F}^{(m)}(\textbf{V} )= \partial_{i\chi}^m\mathcal{F}(\chi,\textbf{V})|_{\chi=0}$ and
\begin{equation}
\mathcal{F}^{(1)}  =  \frac{ q \beta e^{(V_L+V_R)/V_s}}{4 V_s\left(e^{V_L/V_s}\!+\!e^{V_R/V_s}\right)^2 \cosh^2 \left(\beta\Delta E/2\right)}
\end{equation}
as shown in Fig.~\ref{fig:CurrentFano}a. For the variance $\langle\!\langle n^2 \rangle\!\rangle/\mathcal{N} \simeq 2k_BT\int_0^1 ds G(s)/q^2f\pm \int\!\!\!\int_\mathcal{S} d V_L d V_R \mathcal{F}^{(2)}(\textbf{V} )$, we have
\begin{equation}
    \mathcal{F}^{(2)} =     \frac{q\beta e^{\left(V_R+V_L\right)/V_s}\left(e^{ V_R/V_s}\!-\!e^{V_L/V_s}\right) \sinh^4(\beta \Delta E/2) }{32 V_s\left(e^{V_R/V_s}\!+\!e^{V_L/V_s}\right)^3 \sinh^3(\beta \Delta E)}
\end{equation}
as shown in Fig.~\ref{fig:CurrentFano}b. We can now position our contour, so that the pumped charge is maximized, and the noise is minimized. To this end, we exploit the symmetry $\mathcal{F}^{(j)}(V_L,V_R)=\mathcal{F}^{(j)}(-V_C-V_L,-V_C-V_R)$, $j=1,2$, about the point $(-V_C/2,-V_C/2)$ with $V_C=q/2C$, together with the symmetry $\mathcal{F}^{(j)}(V_L,V_R)=(-1)^{j-1}\mathcal{F}^{(j)}(V_R,V_L)$ across the line $V_L=V_R$. Specifically, for a fixed shape of the contour, the contribution to the variance vanishes, if the contour is placed symmetrically across the line $V_L=V_R$. In that case, the noise is due to equilibrium fluctuations only. Moreover, the pumped charge is maximized, if the contour is also symmetric about the point $(-V_C/2,-V_C/2)$.

Figure \ref{fig:CurrentFano} shows the pumped charge and the Fano factor for the driving protocols indicated in the insets together with the Berry curvatures. As in the experiments of Refs.~\cite{Jehl2013,Ono2003}, we consider elliptic contours in the parameter space. Both the red and the blue ellipse minimize the noise, while only the red one also maximizes the pumped charge. In the high-frequency regime, the pumped charge $\langle n \rangle/\mathcal{N} \simeq \mathcal{T} \partial_{i\chi}\varphi^{(1)} (\chi)|_{\chi=0}$ decreases as $1/f^{2}$, since there is no contribution from $\varphi^{(0)}(\chi)$ without a voltage bias. The variance, by contrast, is dominated by thermal fluctuations described by $\varphi^{(0)}(\chi)$. We then have $\langle\!\langle n^2 \rangle\!\rangle/\mathcal{N} \simeq  \mathcal{T} \partial^2_{i\chi}\varphi^{(0)} (\chi)|_{\chi=0}$, implying that the Fano factor is proportional to the frequency. These conclusions are supported by our numerical results in Fig.~\ref{fig:CurrentFano}. At low frequencies, the Fano factor is very small (not visible in the figure) and inversely proportional to the frequency.  Importantly, from our high-frequency expansion, we get a good estimate of the breakdown frequency for which a quantized current can no longer be generated.

\section{Conclusions} We have \revision{employed full counting statistics techniques to optimize the operation of charge pumps. To this end, we have used Floquet theory to evaluate the cumulant generating function for the distribution of pumped charge perturbatively in the frequency or the period of the drive. For the device optimization, we have focused on the average and the variance (noise) of the pumped charge, but higher cumulants, or even the large-deviation statistics, can be obtained along the same lines with little added effort. Our theoretical framework  covers a wide range of driving frequencies, in the adiabatic regime and for fast driving, and it is useful for practical device optimization.} The advances reported here were made possible due to the  progress made in theories of driven systems.  Our work demonstrates that full counting statistics is a powerful tool to optimize charge pumps, and our predictions may be confirmed in future experiments.

\acknowledgements
\revision{We thank V.~Kashcheyevs and T.~Ojanen for useful discussions.} K.~B.~acknowledges support from Academy of Finland (Contract No.~296073). The work was supported by Academy of Finland (projects No.~308515 and 312299). All  authors  are  associated  with the Centre
for Quantum Engineering at Aalto University.


\begin{thebibliography}{57}%
\makeatletter
\providecommand \@ifxundefined [1]{%
 \@ifx{#1\undefined}
}%
\providecommand \@ifnum [1]{%
 \ifnum #1\expandafter \@firstoftwo
 \else \expandafter \@secondoftwo
 \fi
}%
\providecommand \@ifx [1]{%
 \ifx #1\expandafter \@firstoftwo
 \else \expandafter \@secondoftwo
 \fi
}%
\providecommand \natexlab [1]{#1}%
\providecommand \enquote  [1]{``#1''}%
\providecommand \bibnamefont  [1]{#1}%
\providecommand \bibfnamefont [1]{#1}%
\providecommand \citenamefont [1]{#1}%
\providecommand \href@noop [0]{\@secondoftwo}%
\providecommand \href [0]{\begingroup \@sanitize@url \@href}%
\providecommand \@href[1]{\@@startlink{#1}\@@href}%
\providecommand \@@href[1]{\endgroup#1\@@endlink}%
\providecommand \@sanitize@url [0]{\catcode `\\12\catcode `\$12\catcode
  `\&12\catcode `\#12\catcode `\^12\catcode `\_12\catcode `\%12\relax}%
\providecommand \@@startlink[1]{}%
\providecommand \@@endlink[0]{}%
\providecommand \url  [0]{\begingroup\@sanitize@url \@url }%
\providecommand \@url [1]{\endgroup\@href {#1}{\urlprefix }}%
\providecommand \urlprefix  [0]{URL }%
\providecommand \Eprint [0]{\href }%
\providecommand \doibase [0]{http://dx.doi.org/}%
\providecommand \selectlanguage [0]{\@gobble}%
\providecommand \bibinfo  [0]{\@secondoftwo}%
\providecommand \bibfield  [0]{\@secondoftwo}%
\providecommand \translation [1]{[#1]}%
\providecommand \BibitemOpen [0]{}%
\providecommand \bibitemStop [0]{}%
\providecommand \bibitemNoStop [0]{.\EOS\space}%
\providecommand \EOS [0]{\spacefactor3000\relax}%
\providecommand \BibitemShut  [1]{\csname bibitem#1\endcsname}%
\let\auto@bib@innerbib\@empty
\bibitem [{\citenamefont {Odintsov}(1991)}]{Odintsov1991}%
  \BibitemOpen
  \bibfield  {author} {\bibinfo {author} {\bibfnamefont {A.~A.}\ \bibnamefont
  {Odintsov}},\ }\bibfield  {title} {\enquote {\bibinfo {title} {Single
  electron transport in a two‐-dimensional electron gas system with modulated
  barriers: A possible dc current standard},}\ }\href {\doibase
  10.1063/1.104786} {\bibfield  {journal} {\bibinfo  {journal} {Appl. Phys.
  Lett.}\ }\textbf {\bibinfo {volume} {58}},\ \bibinfo {pages} {2697} (\bibinfo
  {year} {1991})}\BibitemShut {NoStop}%
\bibitem [{\citenamefont {Giblin}\ \emph {et~al.}(2012)\citenamefont {Giblin},
  \citenamefont {Kataoka}, \citenamefont {Fletcher}, \citenamefont {See},
  \citenamefont {Janssen}, \citenamefont {Griffiths}, \citenamefont {Jones},
  \citenamefont {Farrer},\ and\ \citenamefont {Ritchie}}]{Giblin2012}%
  \BibitemOpen
  \bibfield  {author} {\bibinfo {author} {\bibfnamefont {S.~P.}\ \bibnamefont
  {Giblin}}, \bibinfo {author} {\bibfnamefont {M.}~\bibnamefont {Kataoka}},
  \bibinfo {author} {\bibfnamefont {J.~D.}\ \bibnamefont {Fletcher}}, \bibinfo
  {author} {\bibfnamefont {P.}~\bibnamefont {See}}, \bibinfo {author}
  {\bibfnamefont {T.~J. B.~M.}\ \bibnamefont {Janssen}}, \bibinfo {author}
  {\bibfnamefont {J.~P.}\ \bibnamefont {Griffiths}}, \bibinfo {author}
  {\bibfnamefont {G.~A.~C.}\ \bibnamefont {Jones}}, \bibinfo {author}
  {\bibfnamefont {I.}~\bibnamefont {Farrer}}, \ and\ \bibinfo {author}
  {\bibfnamefont {D.~A.}\ \bibnamefont {Ritchie}},\ }\bibfield  {title}
  {\enquote {\bibinfo {title} {Towards a quantum representation of the ampere
  using single electron pumps},}\ }\href {http://dx.doi.org/10.1038/ncomms1935}
  {\bibfield  {journal} {\bibinfo  {journal} {Nat. Commun.}\ }\textbf {\bibinfo
  {volume} {3}},\ \bibinfo {pages} {930} (\bibinfo {year} {2012})}\BibitemShut
  {NoStop}%
\bibitem [{\citenamefont {Pekola}\ \emph {et~al.}(2013)\citenamefont {Pekola},
  \citenamefont {Saira}, \citenamefont {Maisi}, \citenamefont {Kemppinen},
  \citenamefont {M\"ott\"onen}, \citenamefont {Pashkin},\ and\ \citenamefont
  {Averin}}]{Pekola2013}%
  \BibitemOpen
  \bibfield  {author} {\bibinfo {author} {\bibfnamefont {J.~P.}\ \bibnamefont
  {Pekola}}, \bibinfo {author} {\bibfnamefont {O.-P.}\ \bibnamefont {Saira}},
  \bibinfo {author} {\bibfnamefont {V.~F.}\ \bibnamefont {Maisi}}, \bibinfo
  {author} {\bibfnamefont {A.}~\bibnamefont {Kemppinen}}, \bibinfo {author}
  {\bibfnamefont {M.}~\bibnamefont {M\"ott\"onen}}, \bibinfo {author}
  {\bibfnamefont {Yu.~A.}\ \bibnamefont {Pashkin}}, \ and\ \bibinfo {author}
  {\bibfnamefont {D.~V.}\ \bibnamefont {Averin}},\ }\bibfield  {title}
  {\enquote {\bibinfo {title} {Single-electron current sources: {T}oward a
  refined definition of the ampere},}\ }\href {\doibase
  10.1103/RevModPhys.85.1421} {\bibfield  {journal} {\bibinfo  {journal} {Rev.
  Mod. Phys.}\ }\textbf {\bibinfo {volume} {85}},\ \bibinfo {pages} {1472}
  (\bibinfo {year} {2013})}\BibitemShut {NoStop}%
\bibitem [{\citenamefont {Kouwenhoven}\ \emph {et~al.}(1991)\citenamefont
  {Kouwenhoven}, \citenamefont {Johnson}, \citenamefont {van~der Vaart},
  \citenamefont {Harmans},\ and\ \citenamefont {Foxon}}]{Kouwenhoven1991}%
  \BibitemOpen
  \bibfield  {author} {\bibinfo {author} {\bibfnamefont {L.~P.}\ \bibnamefont
  {Kouwenhoven}}, \bibinfo {author} {\bibfnamefont {A.~T.}\ \bibnamefont
  {Johnson}}, \bibinfo {author} {\bibfnamefont {N.~C.}\ \bibnamefont {van~der
  Vaart}}, \bibinfo {author} {\bibfnamefont {C.~J. P.~M.}\ \bibnamefont
  {Harmans}}, \ and\ \bibinfo {author} {\bibfnamefont {C.~T.}\ \bibnamefont
  {Foxon}},\ }\bibfield  {title} {\enquote {\bibinfo {title} {{Q}uantized
  {C}urrent in a {Q}uantum-{D}ot {T}urnstile {U}sing {O}scillating {T}unnel
  {B}arriers},}\ }\href {\doibase 10.1103/PhysRevLett.67.1626} {\bibfield
  {journal} {\bibinfo  {journal} {Phys. Rev. Lett.}\ }\textbf {\bibinfo
  {volume} {67}},\ \bibinfo {pages} {1626} (\bibinfo {year}
  {1991})}\BibitemShut {NoStop}%
\bibitem [{\citenamefont {Pothier}\ \emph {et~al.}(1992)\citenamefont
  {Pothier}, \citenamefont {Lafarge}, \citenamefont {Urbina}, \citenamefont
  {Esteve},\ and\ \citenamefont {Devoret}}]{Pothier1992}%
  \BibitemOpen
  \bibfield  {author} {\bibinfo {author} {\bibfnamefont {H.}~\bibnamefont
  {Pothier}}, \bibinfo {author} {\bibfnamefont {P.}~\bibnamefont {Lafarge}},
  \bibinfo {author} {\bibfnamefont {C.}~\bibnamefont {Urbina}}, \bibinfo
  {author} {\bibfnamefont {D.}~\bibnamefont {Esteve}}, \ and\ \bibinfo {author}
  {\bibfnamefont {M.~H.}\ \bibnamefont {Devoret}},\ }\bibfield  {title}
  {\enquote {\bibinfo {title} {{S}ingle-{E}lectron {P}ump {B}ased on {C}harging
  {E}ffects},}\ }\href {http://stacks.iop.org/0295-5075/17/i=3/a=011}
  {\bibfield  {journal} {\bibinfo  {journal} {Europhys. Lett.}\ }\textbf
  {\bibinfo {volume} {17}},\ \bibinfo {pages} {249} (\bibinfo {year}
  {1992})}\BibitemShut {NoStop}%
\bibitem [{\citenamefont {Keller}\ \emph {et~al.}(1996)\citenamefont {Keller},
  \citenamefont {Martinis}, \citenamefont {Zimmerman},\ and\ \citenamefont
  {Steinbach}}]{Keller1996}%
  \BibitemOpen
  \bibfield  {author} {\bibinfo {author} {\bibfnamefont {M.~W.}\ \bibnamefont
  {Keller}}, \bibinfo {author} {\bibfnamefont {J.~M.}\ \bibnamefont
  {Martinis}}, \bibinfo {author} {\bibfnamefont {N.~M.}\ \bibnamefont
  {Zimmerman}}, \ and\ \bibinfo {author} {\bibfnamefont {A.~H.}\ \bibnamefont
  {Steinbach}},\ }\bibfield  {title} {\enquote {\bibinfo {title} {Accuracy of
  electron counting using a 7-junction electron pump},}\ }\href {\doibase
  10.1063/1.117492} {\bibfield  {journal} {\bibinfo  {journal} {Appl. Phys.
  Lett.}\ }\textbf {\bibinfo {volume} {69}},\ \bibinfo {pages} {1804} (\bibinfo
  {year} {1996})}\BibitemShut {NoStop}%
\bibitem [{\citenamefont {Ono}\ and\ \citenamefont
  {Takahashi}(2003)}]{Ono2003}%
  \BibitemOpen
  \bibfield  {author} {\bibinfo {author} {\bibfnamefont {Y.}~\bibnamefont
  {Ono}}\ and\ \bibinfo {author} {\bibfnamefont {Y.}~\bibnamefont
  {Takahashi}},\ }\bibfield  {title} {\enquote {\bibinfo {title} {Electron pump
  by a combined single-electron/field-effect-transistor~structure},}\ }\href
  {\doibase 10.1063/1.1556558} {\bibfield  {journal} {\bibinfo  {journal}
  {Appl. Phys. Lett.}\ }\textbf {\bibinfo {volume} {82}},\ \bibinfo {pages}
  {1223} (\bibinfo {year} {2003})}\BibitemShut {NoStop}%
\bibitem [{\citenamefont {Robinson}\ and\ \citenamefont
  {Talyanskii}(2005)}]{Robinson2005}%
  \BibitemOpen
  \bibfield  {author} {\bibinfo {author} {\bibfnamefont {A.~M.}\ \bibnamefont
  {Robinson}}\ and\ \bibinfo {author} {\bibfnamefont {V.~I.}\ \bibnamefont
  {Talyanskii}},\ }\bibfield  {title} {\enquote {\bibinfo {title} {Shot {N}oise
  in the {C}urrent of a {S}urface {A}coustic-{W}ave-{D}riven
  {S}ingle-{E}lectron {P}ump},}\ }\href {\doibase
  10.1103/PhysRevLett.95.247202} {\bibfield  {journal} {\bibinfo  {journal}
  {Phys. Rev. Lett.}\ }\textbf {\bibinfo {volume} {95}},\ \bibinfo {pages}
  {247202} (\bibinfo {year} {2005})}\BibitemShut {NoStop}%
\bibitem [{\citenamefont {Pekola}\ \emph {et~al.}(2007)\citenamefont {Pekola},
  \citenamefont {Vartiainen}, \citenamefont {M{\"o}tt{\"o}nen}, \citenamefont
  {Saira}, \citenamefont {Meschke},\ and\ \citenamefont {Averin}}]{Pekola2007}%
  \BibitemOpen
  \bibfield  {author} {\bibinfo {author} {\bibfnamefont {J.~P.}\ \bibnamefont
  {Pekola}}, \bibinfo {author} {\bibfnamefont {J.~J.}\ \bibnamefont
  {Vartiainen}}, \bibinfo {author} {\bibfnamefont {M.}~\bibnamefont
  {M{\"o}tt{\"o}nen}}, \bibinfo {author} {\bibfnamefont {O.-P.}\ \bibnamefont
  {Saira}}, \bibinfo {author} {\bibfnamefont {M.}~\bibnamefont {Meschke}}, \
  and\ \bibinfo {author} {\bibfnamefont {D.~V.}\ \bibnamefont {Averin}},\
  }\bibfield  {title} {\enquote {\bibinfo {title} {Hybrid single-electron
  transistor as a source of quantized electric current},}\ }\href
  {http://dx.doi.org/10.1038/nphys808} {\bibfield  {journal} {\bibinfo
  {journal} {Nat. Phys.}\ }\textbf {\bibinfo {volume} {4}},\ \bibinfo {pages}
  {120} (\bibinfo {year} {2007})}\BibitemShut {NoStop}%
\bibitem [{\citenamefont {Fujiwara}\ \emph {et~al.}(2008)\citenamefont
  {Fujiwara}, \citenamefont {Nishiguchi},\ and\ \citenamefont
  {Ono}}]{Fujiwara2008}%
  \BibitemOpen
  \bibfield  {author} {\bibinfo {author} {\bibfnamefont {A.}~\bibnamefont
  {Fujiwara}}, \bibinfo {author} {\bibfnamefont {K.}~\bibnamefont
  {Nishiguchi}}, \ and\ \bibinfo {author} {\bibfnamefont {Y.}~\bibnamefont
  {Ono}},\ }\bibfield  {title} {\enquote {\bibinfo {title} {Nanoampere charge
  pump by single-electron ratchet using silicon nanowire
  metal-oxide-semiconductor field-effect transistor},}\ }\href {\doibase
  10.1063/1.2837544} {\bibfield  {journal} {\bibinfo  {journal} {Appl. Phys.
  Lett.}\ }\textbf {\bibinfo {volume} {92}},\ \bibinfo {pages} {042102}
  (\bibinfo {year} {2008})}\BibitemShut {NoStop}%
\bibitem [{\citenamefont {Jehl}\ \emph {et~al.}(2013)\citenamefont {Jehl},
  \citenamefont {Voisin}, \citenamefont {Charron}, \citenamefont {Clapera},
  \citenamefont {Ray}, \citenamefont {Roche}, \citenamefont {Sanquer},
  \citenamefont {Djordjevic}, \citenamefont {Devoille}, \citenamefont
  {Wacquez},\ and\ \citenamefont {Vinet}}]{Jehl2013}%
  \BibitemOpen
  \bibfield  {author} {\bibinfo {author} {\bibfnamefont {X.}~\bibnamefont
  {Jehl}}, \bibinfo {author} {\bibfnamefont {B.}~\bibnamefont {Voisin}},
  \bibinfo {author} {\bibfnamefont {T.}~\bibnamefont {Charron}}, \bibinfo
  {author} {\bibfnamefont {P.}~\bibnamefont {Clapera}}, \bibinfo {author}
  {\bibfnamefont {S.}~\bibnamefont {Ray}}, \bibinfo {author} {\bibfnamefont
  {B.}~\bibnamefont {Roche}}, \bibinfo {author} {\bibfnamefont
  {M.}~\bibnamefont {Sanquer}}, \bibinfo {author} {\bibfnamefont
  {S.}~\bibnamefont {Djordjevic}}, \bibinfo {author} {\bibfnamefont
  {L.}~\bibnamefont {Devoille}}, \bibinfo {author} {\bibfnamefont
  {R.}~\bibnamefont {Wacquez}}, \ and\ \bibinfo {author} {\bibfnamefont
  {M.}~\bibnamefont {Vinet}},\ }\bibfield  {title} {\enquote {\bibinfo {title}
  {{H}ybrid {M}etal-{S}emiconductor {E}lectron {P}ump for {Q}uantum
  {M}etrology},}\ }\href {\doibase 10.1103/PhysRevX.3.021012} {\bibfield
  {journal} {\bibinfo  {journal} {Phys. Rev. X}\ }\textbf {\bibinfo {volume}
  {3}},\ \bibinfo {pages} {021012} (\bibinfo {year} {2013})}\BibitemShut
  {NoStop}%
\bibitem [{\citenamefont {Yamahata}\ \emph {et~al.}(2014)\citenamefont
  {Yamahata}, \citenamefont {Nishiguchi},\ and\ \citenamefont
  {Fujiwara}}]{Yamahata2014}%
  \BibitemOpen
  \bibfield  {author} {\bibinfo {author} {\bibfnamefont {G.}~\bibnamefont
  {Yamahata}}, \bibinfo {author} {\bibfnamefont {K.}~\bibnamefont
  {Nishiguchi}}, \ and\ \bibinfo {author} {\bibfnamefont {A.}~\bibnamefont
  {Fujiwara}},\ }\bibfield  {title} {\enquote {\bibinfo {title} {Gigahertz
  single-trap electron pumps in silicon},}\ }\href
  {http://dx.doi.org/10.1038/ncomms6038} {\bibfield  {journal} {\bibinfo
  {journal} {Nat. Commun.}\ }\textbf {\bibinfo {volume} {5}},\ \bibinfo {pages}
  {5038} (\bibinfo {year} {2014})}\BibitemShut {NoStop}%
\bibitem [{\citenamefont {Rossi}\ \emph {et~al.}(2014)\citenamefont {Rossi},
  \citenamefont {Tanttu}, \citenamefont {Tan}, \citenamefont {Iisakka},
  \citenamefont {Zhao}, \citenamefont {Chan}, \citenamefont {Tettamanzi},
  \citenamefont {Rogge}, \citenamefont {Dzurak},\ and\ \citenamefont
  {M{\"o}tt{\"o}nen}}]{Rossi2014}%
  \BibitemOpen
  \bibfield  {author} {\bibinfo {author} {\bibfnamefont {A.}~\bibnamefont
  {Rossi}}, \bibinfo {author} {\bibfnamefont {T.}~\bibnamefont {Tanttu}},
  \bibinfo {author} {\bibfnamefont {K.~Y.}\ \bibnamefont {Tan}}, \bibinfo
  {author} {\bibfnamefont {I.}~\bibnamefont {Iisakka}}, \bibinfo {author}
  {\bibfnamefont {R.}~\bibnamefont {Zhao}}, \bibinfo {author} {\bibfnamefont
  {K.~W.}\ \bibnamefont {Chan}}, \bibinfo {author} {\bibfnamefont {G.~C.}\
  \bibnamefont {Tettamanzi}}, \bibinfo {author} {\bibfnamefont
  {S.}~\bibnamefont {Rogge}}, \bibinfo {author} {\bibfnamefont {A.~S.}\
  \bibnamefont {Dzurak}}, \ and\ \bibinfo {author} {\bibfnamefont
  {M.}~\bibnamefont {M{\"o}tt{\"o}nen}},\ }\bibfield  {title} {\enquote
  {\bibinfo {title} {An {A}ccurate {S}ingle-{E}lectron {P}ump {B}ased on a
  {H}ighly {T}unable {S}ilicon {Q}uantum {D}ot},}\ }\href {\doibase
  10.1021/nl500927q} {\bibfield  {journal} {\bibinfo  {journal} {Nano Lett.}\
  }\textbf {\bibinfo {volume} {14}},\ \bibinfo {pages} {3411} (\bibinfo {year}
  {2014})}\BibitemShut {NoStop}%
\bibitem [{\citenamefont {Connolly}\ \emph {et~al.}(2013)\citenamefont
  {Connolly}, \citenamefont {Chiu}, \citenamefont {Giblin}, \citenamefont
  {Kataoka}, \citenamefont {Fletcher}, \citenamefont {Chua}, \citenamefont
  {Griffiths}, \citenamefont {Jones}, \citenamefont {Fal'ko}, \citenamefont
  {Smith},\ and\ \citenamefont {Janssen}}]{Connolly2013}%
  \BibitemOpen
  \bibfield  {author} {\bibinfo {author} {\bibfnamefont {M.~R.}\ \bibnamefont
  {Connolly}}, \bibinfo {author} {\bibfnamefont {K.~L.}\ \bibnamefont {Chiu}},
  \bibinfo {author} {\bibfnamefont {S.~P.}\ \bibnamefont {Giblin}}, \bibinfo
  {author} {\bibfnamefont {M.}~\bibnamefont {Kataoka}}, \bibinfo {author}
  {\bibfnamefont {J.~D.}\ \bibnamefont {Fletcher}}, \bibinfo {author}
  {\bibfnamefont {C.}~\bibnamefont {Chua}}, \bibinfo {author} {\bibfnamefont
  {J.~P.}\ \bibnamefont {Griffiths}}, \bibinfo {author} {\bibfnamefont
  {G.~A.~C.}\ \bibnamefont {Jones}}, \bibinfo {author} {\bibfnamefont {V.~I.}\
  \bibnamefont {Fal'ko}}, \bibinfo {author} {\bibfnamefont {C.~G.}\
  \bibnamefont {Smith}}, \ and\ \bibinfo {author} {\bibfnamefont {T.~J. B.~M.}\
  \bibnamefont {Janssen}},\ }\bibfield  {title} {\enquote {\bibinfo {title}
  {Gigahertz quantized charge pumping in graphene quantum dots},}\ }\href
  {http://dx.doi.org/10.1038/nnano.2013.73} {\bibfield  {journal} {\bibinfo
  {journal} {Nat. Nanotech.}\ }\textbf {\bibinfo {volume} {8}},\ \bibinfo
  {pages} {417} (\bibinfo {year} {2013})}\BibitemShut {NoStop}%
\bibitem [{\citenamefont {Blumenthal}\ \emph {et~al.}(2007)\citenamefont
  {Blumenthal}, \citenamefont {Kaestner}, \citenamefont {Li}, \citenamefont
  {Giblin}, \citenamefont {Janssen}, \citenamefont {Pepper}, \citenamefont
  {Anderson}, \citenamefont {Jones},\ and\ \citenamefont
  {Ritchie}}]{Blumenthal2007}%
  \BibitemOpen
  \bibfield  {author} {\bibinfo {author} {\bibfnamefont {M.~D.}\ \bibnamefont
  {Blumenthal}}, \bibinfo {author} {\bibfnamefont {B.}~\bibnamefont
  {Kaestner}}, \bibinfo {author} {\bibfnamefont {L.}~\bibnamefont {Li}},
  \bibinfo {author} {\bibfnamefont {S.}~\bibnamefont {Giblin}}, \bibinfo
  {author} {\bibfnamefont {T.~J. B.~M.}\ \bibnamefont {Janssen}}, \bibinfo
  {author} {\bibfnamefont {M.}~\bibnamefont {Pepper}}, \bibinfo {author}
  {\bibfnamefont {D.}~\bibnamefont {Anderson}}, \bibinfo {author}
  {\bibfnamefont {G.}~\bibnamefont {Jones}}, \ and\ \bibinfo {author}
  {\bibfnamefont {D.~A.}\ \bibnamefont {Ritchie}},\ }\bibfield  {title}
  {\enquote {\bibinfo {title} {Gigahertz quantized charge pumping},}\ }\href
  {\doibase 10.1038/nphys582} {\bibfield  {journal} {\bibinfo  {journal} {Nat.
  Phys.}\ }\textbf {\bibinfo {volume} {3}},\ \bibinfo {pages} {347} (\bibinfo
  {year} {2007})}\BibitemShut {NoStop}%
\bibitem [{\citenamefont {Kaestner}\ \emph {et~al.}(2008)\citenamefont
  {Kaestner}, \citenamefont {Kashcheyevs}, \citenamefont {Amakawa},
  \citenamefont {Blumenthal}, \citenamefont {Li}, \citenamefont {Janssen},
  \citenamefont {Hein}, \citenamefont {Pierz}, \citenamefont {Weimann},
  \citenamefont {Siegner},\ and\ \citenamefont {Schumacher}}]{SlavaBernd2008}%
  \BibitemOpen
  \bibfield  {author} {\bibinfo {author} {\bibfnamefont {B.}~\bibnamefont
  {Kaestner}}, \bibinfo {author} {\bibfnamefont {V.}~\bibnamefont
  {Kashcheyevs}}, \bibinfo {author} {\bibfnamefont {S.}~\bibnamefont
  {Amakawa}}, \bibinfo {author} {\bibfnamefont {M.~D.}\ \bibnamefont
  {Blumenthal}}, \bibinfo {author} {\bibfnamefont {L.}~\bibnamefont {Li}},
  \bibinfo {author} {\bibfnamefont {T.~J. B.~M.}\ \bibnamefont {Janssen}},
  \bibinfo {author} {\bibfnamefont {G.}~\bibnamefont {Hein}}, \bibinfo {author}
  {\bibfnamefont {K.}~\bibnamefont {Pierz}}, \bibinfo {author} {\bibfnamefont
  {T.}~\bibnamefont {Weimann}}, \bibinfo {author} {\bibfnamefont
  {U.}~\bibnamefont {Siegner}}, \ and\ \bibinfo {author} {\bibfnamefont
  {H.~W.}\ \bibnamefont {Schumacher}},\ }\bibfield  {title} {\enquote {\bibinfo
  {title} {Single-parameter nonadiabatic quantized charge pumping},}\ }\href
  {\doibase 10.1103/PhysRevB.77.153301} {\bibfield  {journal} {\bibinfo
  {journal} {Phys. Rev. B}\ }\textbf {\bibinfo {volume} {77}},\ \bibinfo
  {pages} {153301} (\bibinfo {year} {2008})}\BibitemShut {NoStop}%
\bibitem [{\citenamefont {Giblin}\ \emph {et~al.}(2010)\citenamefont {Giblin},
  \citenamefont {Wright}, \citenamefont {Fletcher}, \citenamefont {Kataoka},
  \citenamefont {Pepper}, \citenamefont {Janssen}, \citenamefont {Ritchie},
  \citenamefont {Nicoll}, \citenamefont {Anderson},\ and\ \citenamefont
  {Jones}}]{Giblin2010}%
  \BibitemOpen
  \bibfield  {author} {\bibinfo {author} {\bibfnamefont {S.~P.}\ \bibnamefont
  {Giblin}}, \bibinfo {author} {\bibfnamefont {S.~J.}\ \bibnamefont {Wright}},
  \bibinfo {author} {\bibfnamefont {J.~D.}\ \bibnamefont {Fletcher}}, \bibinfo
  {author} {\bibfnamefont {M.}~\bibnamefont {Kataoka}}, \bibinfo {author}
  {\bibfnamefont {M.}~\bibnamefont {Pepper}}, \bibinfo {author} {\bibfnamefont
  {T.~J. B.~M.}\ \bibnamefont {Janssen}}, \bibinfo {author} {\bibfnamefont
  {D.~A.}\ \bibnamefont {Ritchie}}, \bibinfo {author} {\bibfnamefont {C.~A.}\
  \bibnamefont {Nicoll}}, \bibinfo {author} {\bibfnamefont {D.}~\bibnamefont
  {Anderson}}, \ and\ \bibinfo {author} {\bibfnamefont {G~A.~C.}\ \bibnamefont
  {Jones}},\ }\bibfield  {title} {\enquote {\bibinfo {title} {An accurate
  high-speed single-electron quantum dot pump},}\ }\href
  {http://stacks.iop.org/1367-2630/12/i=7/a=073013} {\bibfield  {journal}
  {\bibinfo  {journal} {New J. Phys.}\ }\textbf {\bibinfo {volume} {12}},\
  \bibinfo {pages} {073013} (\bibinfo {year} {2010})}\BibitemShut {NoStop}%
\bibitem [{\citenamefont {Kataoka}\ \emph {et~al.}(2011)\citenamefont
  {Kataoka}, \citenamefont {Fletcher}, \citenamefont {See}, \citenamefont
  {Giblin}, \citenamefont {Janssen}, \citenamefont {Griffiths}, \citenamefont
  {Jones}, \citenamefont {Farrer},\ and\ \citenamefont
  {Ritchie}}]{Kataoka2011}%
  \BibitemOpen
  \bibfield  {author} {\bibinfo {author} {\bibfnamefont {M.}~\bibnamefont
  {Kataoka}}, \bibinfo {author} {\bibfnamefont {J.~D.}\ \bibnamefont
  {Fletcher}}, \bibinfo {author} {\bibfnamefont {P.}~\bibnamefont {See}},
  \bibinfo {author} {\bibfnamefont {S.~P.}\ \bibnamefont {Giblin}}, \bibinfo
  {author} {\bibfnamefont {T.~J. B.~M.}\ \bibnamefont {Janssen}}, \bibinfo
  {author} {\bibfnamefont {J.~P.}\ \bibnamefont {Griffiths}}, \bibinfo {author}
  {\bibfnamefont {G.~A.~C.}\ \bibnamefont {Jones}}, \bibinfo {author}
  {\bibfnamefont {I.}~\bibnamefont {Farrer}}, \ and\ \bibinfo {author}
  {\bibfnamefont {D.~A.}\ \bibnamefont {Ritchie}},\ }\bibfield  {title}
  {\enquote {\bibinfo {title} {Tunable {N}onadiabatic {E}xcitation in a
  {S}ingle-{E}lectron {Q}uantum {D}ot},}\ }\href {\doibase
  10.1103/PhysRevLett.106.126801} {\bibfield  {journal} {\bibinfo  {journal}
  {Phys. Rev. Lett.}\ }\textbf {\bibinfo {volume} {106}},\ \bibinfo {pages}
  {126801} (\bibinfo {year} {2011})}\BibitemShut {NoStop}%
\bibitem [{\citenamefont {Fricke}\ \emph {et~al.}(2014)\citenamefont {Fricke},
  \citenamefont {Wulf}, \citenamefont {Kaestner}, \citenamefont {Hohls},
  \citenamefont {Mirovsky}, \citenamefont {Mackrodt}, \citenamefont {Dolata},
  \citenamefont {Weimann}, \citenamefont {Pierz}, \citenamefont {Siegner},\
  and\ \citenamefont {Schumacher}}]{Fricke2014}%
  \BibitemOpen
  \bibfield  {author} {\bibinfo {author} {\bibfnamefont {L.}~\bibnamefont
  {Fricke}}, \bibinfo {author} {\bibfnamefont {M.}~\bibnamefont {Wulf}},
  \bibinfo {author} {\bibfnamefont {B.}~\bibnamefont {Kaestner}}, \bibinfo
  {author} {\bibfnamefont {F.}~\bibnamefont {Hohls}}, \bibinfo {author}
  {\bibfnamefont {P.}~\bibnamefont {Mirovsky}}, \bibinfo {author}
  {\bibfnamefont {B.}~\bibnamefont {Mackrodt}}, \bibinfo {author}
  {\bibfnamefont {R.}~\bibnamefont {Dolata}}, \bibinfo {author} {\bibfnamefont
  {T.}~\bibnamefont {Weimann}}, \bibinfo {author} {\bibfnamefont
  {K.}~\bibnamefont {Pierz}}, \bibinfo {author} {\bibfnamefont
  {U.}~\bibnamefont {Siegner}}, \ and\ \bibinfo {author} {\bibfnamefont
  {H.~W.}\ \bibnamefont {Schumacher}},\ }\bibfield  {title} {\enquote {\bibinfo
  {title} {Self-{R}eferenced {S}ingle-{E}lectron {Q}uantized {C}urrent
  {S}ource},}\ }\href {\doibase 10.1103/PhysRevLett.112.226803} {\bibfield
  {journal} {\bibinfo  {journal} {Phys. Rev. Lett.}\ }\textbf {\bibinfo
  {volume} {112}},\ \bibinfo {pages} {226803} (\bibinfo {year}
  {2014})}\BibitemShut {NoStop}%
\bibitem [{\citenamefont {Ubbelohde}\ \emph {et~al.}(2015)\citenamefont
  {Ubbelohde}, \citenamefont {Hohls}, \citenamefont {Kashcheyevs},
  \citenamefont {Wagner}, \citenamefont {Fricke}, \citenamefont {K\"{a}stner},
  \citenamefont {Pierz}, \citenamefont {Schumacher},\ and\ \citenamefont
  {Haug}}]{Ubbelohde2014}%
  \BibitemOpen
  \bibfield  {author} {\bibinfo {author} {\bibfnamefont {N.}~\bibnamefont
  {Ubbelohde}}, \bibinfo {author} {\bibfnamefont {F.}~\bibnamefont {Hohls}},
  \bibinfo {author} {\bibfnamefont {V.}~\bibnamefont {Kashcheyevs}}, \bibinfo
  {author} {\bibfnamefont {T.}~\bibnamefont {Wagner}}, \bibinfo {author}
  {\bibfnamefont {L.}~\bibnamefont {Fricke}}, \bibinfo {author} {\bibfnamefont
  {B.}~\bibnamefont {K\"{a}stner}}, \bibinfo {author} {\bibfnamefont
  {K.}~\bibnamefont {Pierz}}, \bibinfo {author} {\bibfnamefont {H.~W.}\
  \bibnamefont {Schumacher}}, \ and\ \bibinfo {author} {\bibfnamefont {R.~J.}\
  \bibnamefont {Haug}},\ }\bibfield  {title} {\enquote {\bibinfo {title}
  {Partitioning of on-demand electron pairs},}\ }\href {\doibase
  10.1038/nnano.2014.275} {\bibfield  {journal} {\bibinfo  {journal} {Nat.
  Nanotech.}\ }\textbf {\bibinfo {volume} {10}},\ \bibinfo {pages} {46}
  (\bibinfo {year} {2015})}\BibitemShut {NoStop}%
\bibitem [{\citenamefont {Kaestner}\ and\ \citenamefont
  {Kashcheyevs}(2015)}]{KaestnerKashcheyevs2015}%
  \BibitemOpen
  \bibfield  {author} {\bibinfo {author} {\bibfnamefont {B.}~\bibnamefont
  {Kaestner}}\ and\ \bibinfo {author} {\bibfnamefont {V.}~\bibnamefont
  {Kashcheyevs}},\ }\bibfield  {title} {\enquote {\bibinfo {title}
  {Non-adiabatic quantized charge pumping with tunable-barrier quantum dots: a
  review of current progress},}\ }\href {\doibase
  doi:10.1088/0034-4885/78/10/103901} {\bibfield  {journal} {\bibinfo
  {journal} {Rep. Prog. Phys.}\ }\textbf {\bibinfo {volume} {78}},\ \bibinfo
  {pages} {103901} (\bibinfo {year} {2015})}\BibitemShut {NoStop}%
\bibitem [{\citenamefont {Stein}\ \emph {et~al.}(2015)\citenamefont {Stein},
  \citenamefont {Drung}, \citenamefont {Fricke}, \citenamefont {Scherer},
  \citenamefont {Hohls}, \citenamefont {Leicht}, \citenamefont {G\"otz},
  \citenamefont {Krause}, \citenamefont {Behr}, \citenamefont {Pesel},
  \citenamefont {Pierz}, \citenamefont {Siegner}, \citenamefont {Ahlers},\ and\
  \citenamefont {Schumacher}}]{Stein2015}%
  \BibitemOpen
  \bibfield  {author} {\bibinfo {author} {\bibfnamefont {F.}~\bibnamefont
  {Stein}}, \bibinfo {author} {\bibfnamefont {D.}~\bibnamefont {Drung}},
  \bibinfo {author} {\bibfnamefont {L.}~\bibnamefont {Fricke}}, \bibinfo
  {author} {\bibfnamefont {H.}~\bibnamefont {Scherer}}, \bibinfo {author}
  {\bibfnamefont {F.}~\bibnamefont {Hohls}}, \bibinfo {author} {\bibfnamefont
  {C.}~\bibnamefont {Leicht}}, \bibinfo {author} {\bibfnamefont
  {M.}~\bibnamefont {G\"otz}}, \bibinfo {author} {\bibfnamefont
  {C.}~\bibnamefont {Krause}}, \bibinfo {author} {\bibfnamefont
  {R.}~\bibnamefont {Behr}}, \bibinfo {author} {\bibfnamefont {E.}~\bibnamefont
  {Pesel}}, \bibinfo {author} {\bibfnamefont {K.}~\bibnamefont {Pierz}},
  \bibinfo {author} {\bibfnamefont {U.}~\bibnamefont {Siegner}}, \bibinfo
  {author} {\bibfnamefont {F.~J.}\ \bibnamefont {Ahlers}}, \ and\ \bibinfo
  {author} {\bibfnamefont {H.~W.}\ \bibnamefont {Schumacher}},\ }\bibfield
  {title} {\enquote {\bibinfo {title} {Validation of a quantized-current source
  with 0.2 ppm uncertainty},}\ }\href {\doibase 10.1063/1.4930142} {\bibfield
  {journal} {\bibinfo  {journal} {Appl. Phys. Lett.}\ }\textbf {\bibinfo
  {volume} {107}},\ \bibinfo {pages} {103501} (\bibinfo {year}
  {2015})}\BibitemShut {NoStop}%
\bibitem [{\citenamefont {Yamahata}\ \emph {et~al.}(2016)\citenamefont
  {Yamahata}, \citenamefont {Giblin}, \citenamefont {Kataoka}, \citenamefont
  {Karasawa},\ and\ \citenamefont {Fujiwara}}]{Yamahata2016}%
  \BibitemOpen
  \bibfield  {author} {\bibinfo {author} {\bibfnamefont {G.}~\bibnamefont
  {Yamahata}}, \bibinfo {author} {\bibfnamefont {S.~P.}\ \bibnamefont
  {Giblin}}, \bibinfo {author} {\bibfnamefont {M.}~\bibnamefont {Kataoka}},
  \bibinfo {author} {\bibfnamefont {T.}~\bibnamefont {Karasawa}}, \ and\
  \bibinfo {author} {\bibfnamefont {A.}~\bibnamefont {Fujiwara}},\ }\bibfield
  {title} {\enquote {\bibinfo {title} {Gigahertz single-electron pumping in
  silicon with an accuracy better than 9.2 parts in $10^7$},}\ }\href {\doibase
  10.1063/1.4953872} {\bibfield  {journal} {\bibinfo  {journal} {Appl. Phys.
  Lett.}\ }\textbf {\bibinfo {volume} {109}},\ \bibinfo {pages} {013101}
  (\bibinfo {year} {2016})}\BibitemShut {NoStop}%
\bibitem [{\citenamefont {Ahn}\ \emph {et~al.}(2017)\citenamefont {Ahn},
  \citenamefont {Hong}, \citenamefont {Ghee}, \citenamefont {Chung},
  \citenamefont {Hong}, \citenamefont {Bae},\ and\ \citenamefont
  {Kim}}]{Ahn2017}%
  \BibitemOpen
  \bibfield  {author} {\bibinfo {author} {\bibfnamefont {Y.-H.}\ \bibnamefont
  {Ahn}}, \bibinfo {author} {\bibfnamefont {C.}~\bibnamefont {Hong}}, \bibinfo
  {author} {\bibfnamefont {Y.}~\bibnamefont {Ghee}}, \bibinfo {author}
  {\bibfnamefont {Y.}~\bibnamefont {Chung}}, \bibinfo {author} {\bibfnamefont
  {Y.-P.}\ \bibnamefont {Hong}}, \bibinfo {author} {\bibfnamefont {M.-H.}\
  \bibnamefont {Bae}}, \ and\ \bibinfo {author} {\bibfnamefont
  {N.}~\bibnamefont {Kim}},\ }\bibfield  {title} {\enquote {\bibinfo {title}
  {Upper frequency limit depending on potential shape in a {QD}-based single
  electron pump},}\ }\href {\doibase 10.1063/1.5000319} {\bibfield  {journal}
  {\bibinfo  {journal} {J. Appl. Phys.}\ }\textbf {\bibinfo {volume} {122}},\
  \bibinfo {pages} {194502} (\bibinfo {year} {2017})}\BibitemShut {NoStop}%
\bibitem [{\citenamefont {Zhao}\ \emph {et~al.}(2017)\citenamefont {Zhao},
  \citenamefont {Rossi}, \citenamefont {Giblin}, \citenamefont {Fletcher},
  \citenamefont {Hudson}, \citenamefont {M\"ott\"onen}, \citenamefont
  {Kataoka},\ and\ \citenamefont {Dzurak}}]{Zhao2017}%
  \BibitemOpen
  \bibfield  {author} {\bibinfo {author} {\bibfnamefont {R.}~\bibnamefont
  {Zhao}}, \bibinfo {author} {\bibfnamefont {A.}~\bibnamefont {Rossi}},
  \bibinfo {author} {\bibfnamefont {S.~P.}\ \bibnamefont {Giblin}}, \bibinfo
  {author} {\bibfnamefont {J.~D.}\ \bibnamefont {Fletcher}}, \bibinfo {author}
  {\bibfnamefont {F.~E.}\ \bibnamefont {Hudson}}, \bibinfo {author}
  {\bibfnamefont {M.}~\bibnamefont {M\"ott\"onen}}, \bibinfo {author}
  {\bibfnamefont {M.}~\bibnamefont {Kataoka}}, \ and\ \bibinfo {author}
  {\bibfnamefont {A.~S.}\ \bibnamefont {Dzurak}},\ }\bibfield  {title}
  {\enquote {\bibinfo {title} {Thermal-{E}rror {R}egime in {H}igh-{A}ccuracy
  {G}igahertz {S}ingle-{E}lectron {P}umping},}\ }\href {\doibase
  10.1103/PhysRevApplied.8.044021} {\bibfield  {journal} {\bibinfo  {journal}
  {Phys. Rev. Applied}\ }\textbf {\bibinfo {volume} {8}},\ \bibinfo {pages}
  {044021} (\bibinfo {year} {2017})}\BibitemShut {NoStop}%
\bibitem [{\citenamefont {Brun-Picard}\ \emph {et~al.}(2016)\citenamefont
  {Brun-Picard}, \citenamefont {Djordjevic}, \citenamefont {Leprat},
  \citenamefont {Schopfer},\ and\ \citenamefont {Poirier}}]{BrunPicard2016}%
  \BibitemOpen
  \bibfield  {author} {\bibinfo {author} {\bibfnamefont {J.}~\bibnamefont
  {Brun-Picard}}, \bibinfo {author} {\bibfnamefont {S.}~\bibnamefont
  {Djordjevic}}, \bibinfo {author} {\bibfnamefont {D.}~\bibnamefont {Leprat}},
  \bibinfo {author} {\bibfnamefont {F.}~\bibnamefont {Schopfer}}, \ and\
  \bibinfo {author} {\bibfnamefont {W.}~\bibnamefont {Poirier}},\ }\bibfield
  {title} {\enquote {\bibinfo {title} {Practical {Q}uantum {R}ealization of the
  {A}mpere from the {E}lementary {C}harge},}\ }\href {\doibase
  10.1103/PhysRevX.6.041051} {\bibfield  {journal} {\bibinfo  {journal} {Phys.
  Rev. X}\ }\textbf {\bibinfo {volume} {6}},\ \bibinfo {pages} {041051}
  (\bibinfo {year} {2016})}\BibitemShut {NoStop}%
\bibitem [{\citenamefont {Moskalets}\ and\ \citenamefont
  {B\"uttiker}(2002{\natexlab{a}})}]{Moskalets2002}%
  \BibitemOpen
  \bibfield  {author} {\bibinfo {author} {\bibfnamefont {M.}~\bibnamefont
  {Moskalets}}\ and\ \bibinfo {author} {\bibfnamefont {M.}~\bibnamefont
  {B\"uttiker}},\ }\bibfield  {title} {\enquote {\bibinfo {title} {Floquet
  scattering theory of quantum pumps},}\ }\href {\doibase
  10.1103/PhysRevB.66.205320} {\bibfield  {journal} {\bibinfo  {journal} {Phys.
  Rev. B}\ }\textbf {\bibinfo {volume} {66}},\ \bibinfo {pages} {205320}
  (\bibinfo {year} {2002}{\natexlab{a}})}\BibitemShut {NoStop}%
\bibitem [{\citenamefont {Moskalets}\ and\ \citenamefont
  {B\"uttiker}(2002{\natexlab{b}})}]{Moskalets2002b}%
  \BibitemOpen
  \bibfield  {author} {\bibinfo {author} {\bibfnamefont {M.}~\bibnamefont
  {Moskalets}}\ and\ \bibinfo {author} {\bibfnamefont {M.}~\bibnamefont
  {B\"uttiker}},\ }\bibfield  {title} {\enquote {\bibinfo {title} {Dissipation
  and noise in adiabatic quantum pumps},}\ }\href {\doibase
  10.1103/PhysRevB.66.035306} {\bibfield  {journal} {\bibinfo  {journal} {Phys.
  Rev. B}\ }\textbf {\bibinfo {volume} {66}},\ \bibinfo {pages} {035306}
  (\bibinfo {year} {2002}{\natexlab{b}})}\BibitemShut {NoStop}%
\bibitem [{\citenamefont {Brouwer}(1998)}]{Brouwer1998}%
  \BibitemOpen
  \bibfield  {author} {\bibinfo {author} {\bibfnamefont {P.~W.}\ \bibnamefont
  {Brouwer}},\ }\bibfield  {title} {\enquote {\bibinfo {title} {Scattering
  approach to parametric pumping},}\ }\href {\doibase
  10.1103/PhysRevB.58.R10135} {\bibfield  {journal} {\bibinfo  {journal} {Phys.
  Rev. B}\ }\textbf {\bibinfo {volume} {58}},\ \bibinfo {pages} {10135(R)}
  (\bibinfo {year} {1998})}\BibitemShut {NoStop}%
\bibitem [{\citenamefont {Aleiner}\ and\ \citenamefont
  {Andreev}(1998)}]{Aleiner98}%
  \BibitemOpen
  \bibfield  {author} {\bibinfo {author} {\bibfnamefont {I.~L.}\ \bibnamefont
  {Aleiner}}\ and\ \bibinfo {author} {\bibfnamefont {A.~V.}\ \bibnamefont
  {Andreev}},\ }\bibfield  {title} {\enquote {\bibinfo {title} {{A}diabatic
  {C}harge {P}umping in {A}lmost {O}pen {D}ots},}\ }\href {\doibase
  10.1103/PhysRevLett.81.1286} {\bibfield  {journal} {\bibinfo  {journal}
  {Phys. Rev. Lett.}\ }\textbf {\bibinfo {volume} {81}},\ \bibinfo {pages}
  {1286} (\bibinfo {year} {1998})}\BibitemShut {NoStop}%
\bibitem [{\citenamefont {Shutenko}\ \emph {et~al.}(2000)\citenamefont
  {Shutenko}, \citenamefont {Aleiner},\ and\ \citenamefont
  {Altshuler}}]{Shutenko2000}%
  \BibitemOpen
  \bibfield  {author} {\bibinfo {author} {\bibfnamefont {T.~A.}\ \bibnamefont
  {Shutenko}}, \bibinfo {author} {\bibfnamefont {I.~L.}\ \bibnamefont
  {Aleiner}}, \ and\ \bibinfo {author} {\bibfnamefont {B.~L.}\ \bibnamefont
  {Altshuler}},\ }\bibfield  {title} {\enquote {\bibinfo {title} {Mesoscopic
  fluctuations of adiabatic charge pumping in quantum dots},}\ }\href {\doibase
  10.1103/PhysRevB.61.10366} {\bibfield  {journal} {\bibinfo  {journal} {Phys.
  Rev. B}\ }\textbf {\bibinfo {volume} {61}},\ \bibinfo {pages} {10366}
  (\bibinfo {year} {2000})}\BibitemShut {NoStop}%
\bibitem [{\citenamefont {Avron}\ \emph {et~al.}(2000)\citenamefont {Avron},
  \citenamefont {Elgart}, \citenamefont {Graf},\ and\ \citenamefont
  {Sadun}}]{Avron2000}%
  \BibitemOpen
  \bibfield  {author} {\bibinfo {author} {\bibfnamefont {J.~E.}\ \bibnamefont
  {Avron}}, \bibinfo {author} {\bibfnamefont {A.}~\bibnamefont {Elgart}},
  \bibinfo {author} {\bibfnamefont {G.~M.}\ \bibnamefont {Graf}}, \ and\
  \bibinfo {author} {\bibfnamefont {L.}~\bibnamefont {Sadun}},\ }\bibfield
  {title} {\enquote {\bibinfo {title} {Geometry, statistics, and asymptotics of
  quantum pumps},}\ }\href {\doibase 10.1103/PhysRevB.62.R10618} {\bibfield
  {journal} {\bibinfo  {journal} {Phys. Rev. B}\ }\textbf {\bibinfo {volume}
  {62}},\ \bibinfo {pages} {10621(R)} (\bibinfo {year} {2000})}\BibitemShut
  {NoStop}%
\bibitem [{\citenamefont {Makhlin}\ and\ \citenamefont
  {Mirlin}(2001)}]{MakhlinMirlin2001}%
  \BibitemOpen
  \bibfield  {author} {\bibinfo {author} {\bibfnamefont {Yu.}\ \bibnamefont
  {Makhlin}}\ and\ \bibinfo {author} {\bibfnamefont {A.~D.}\ \bibnamefont
  {Mirlin}},\ }\bibfield  {title} {\enquote {\bibinfo {title} {Counting
  {S}tatistics for {A}rbitrary {C}ycles in {Q}uantum {P}umps},}\ }\href
  {\doibase 10.1103/PhysRevLett.87.276803} {\bibfield  {journal} {\bibinfo
  {journal} {Phys. Rev. Lett.}\ }\textbf {\bibinfo {volume} {87}},\ \bibinfo
  {pages} {276803} (\bibinfo {year} {2001})}\BibitemShut {NoStop}%
\bibitem [{\citenamefont {Entin-Wohlman}\ \emph {et~al.}(2002)\citenamefont
  {Entin-Wohlman}, \citenamefont {Aharony},\ and\ \citenamefont
  {Levinson}}]{EntinWohlman2002}%
  \BibitemOpen
  \bibfield  {author} {\bibinfo {author} {\bibfnamefont {O.}~\bibnamefont
  {Entin-Wohlman}}, \bibinfo {author} {\bibfnamefont {A.}~\bibnamefont
  {Aharony}}, \ and\ \bibinfo {author} {\bibfnamefont {Y.}~\bibnamefont
  {Levinson}},\ }\bibfield  {title} {\enquote {\bibinfo {title} {Adiabatic
  transport in nanostructures},}\ }\href {\doibase 10.1103/PhysRevB.65.195411}
  {\bibfield  {journal} {\bibinfo  {journal} {Phys. Rev. B}\ }\textbf {\bibinfo
  {volume} {65}},\ \bibinfo {pages} {195411} (\bibinfo {year}
  {2002})}\BibitemShut {NoStop}%
\bibitem [{\citenamefont {Splettstoesser}\ \emph {et~al.}(2005)\citenamefont
  {Splettstoesser}, \citenamefont {Governale}, \citenamefont {K\"onig},\ and\
  \citenamefont {Fazio}}]{Splettstoesser2005}%
  \BibitemOpen
  \bibfield  {author} {\bibinfo {author} {\bibfnamefont {J.}~\bibnamefont
  {Splettstoesser}}, \bibinfo {author} {\bibfnamefont {M.}~\bibnamefont
  {Governale}}, \bibinfo {author} {\bibfnamefont {J.}~\bibnamefont {K\"onig}},
  \ and\ \bibinfo {author} {\bibfnamefont {R.}~\bibnamefont {Fazio}},\
  }\bibfield  {title} {\enquote {\bibinfo {title} {{A}diabatic {P}umping
  through {I}nteracting {Q}uantum {D}ots},}\ }\href {\doibase
  10.1103/PhysRevLett.95.246803} {\bibfield  {journal} {\bibinfo  {journal}
  {Phys. Rev. Lett.}\ }\textbf {\bibinfo {volume} {95}},\ \bibinfo {pages}
  {246803} (\bibinfo {year} {2005})}\BibitemShut {NoStop}%
\bibitem [{\citenamefont {Kashcheyevs}\ and\ \citenamefont
  {Kaestner}(2010)}]{Kashcheyevs2010}%
  \BibitemOpen
  \bibfield  {author} {\bibinfo {author} {\bibfnamefont {V.}~\bibnamefont
  {Kashcheyevs}}\ and\ \bibinfo {author} {\bibfnamefont {B.}~\bibnamefont
  {Kaestner}},\ }\bibfield  {title} {\enquote {\bibinfo {title} {Universal
  {D}ecay {C}ascade {M}odel for {D}ynamic {Q}uantum {D}ot {I}nitialization},}\
  }\href {\doibase 10.1103/PhysRevLett.104.186805} {\bibfield  {journal}
  {\bibinfo  {journal} {Phys. Rev. Lett.}\ }\textbf {\bibinfo {volume} {104}},\
  \bibinfo {pages} {186805} (\bibinfo {year} {2010})}\BibitemShut {NoStop}%
\bibitem [{\citenamefont {Kashcheyevs}\ and\ \citenamefont
  {Timoshenko}(2012)}]{Kashcheyevs2012}%
  \BibitemOpen
  \bibfield  {author} {\bibinfo {author} {\bibfnamefont {V.}~\bibnamefont
  {Kashcheyevs}}\ and\ \bibinfo {author} {\bibfnamefont {J.}~\bibnamefont
  {Timoshenko}},\ }\bibfield  {title} {\enquote {\bibinfo {title} {Quantum
  {F}luctuations and {C}oherence in {H}igh-{P}recision {S}ingle-{E}lectron
  {C}apture},}\ }\href {\doibase 10.1103/PhysRevLett.109.216801} {\bibfield
  {journal} {\bibinfo  {journal} {Phys. Rev. Lett.}\ }\textbf {\bibinfo
  {volume} {109}},\ \bibinfo {pages} {216801} (\bibinfo {year}
  {2012})}\BibitemShut {NoStop}%
\bibitem [{\citenamefont {Ohkubo}\ and\ \citenamefont
  {Eggel}(2010)}]{Ohkubo2010}%
  \BibitemOpen
  \bibfield  {author} {\bibinfo {author} {\bibfnamefont {J.}~\bibnamefont
  {Ohkubo}}\ and\ \bibinfo {author} {\bibfnamefont {T.}~\bibnamefont {Eggel}},\
  }\bibfield  {title} {\enquote {\bibinfo {title} {A direct numerical method
  for obtaining the counting statistics for stochastic processes},}\ }\href
  {http://stacks.iop.org/1742-5468/2010/i=06/a=P06013} {\bibfield  {journal}
  {\bibinfo  {journal} {J. Stat. Mech.}\ }\textbf {\bibinfo {volume} {2010}},\
  \bibinfo {pages} {P06013} (\bibinfo {year} {2010})}\BibitemShut {NoStop}%
\bibitem [{\citenamefont {Croy}\ and\ \citenamefont
  {Saalmann}(2012)}]{Croy2012}%
  \BibitemOpen
  \bibfield  {author} {\bibinfo {author} {\bibfnamefont {A.}~\bibnamefont
  {Croy}}\ and\ \bibinfo {author} {\bibfnamefont {U.}~\bibnamefont
  {Saalmann}},\ }\bibfield  {title} {\enquote {\bibinfo {title} {Nonadiabatic
  rectification and current reversal in electron pumps},}\ }\href {\doibase
  10.1103/PhysRevB.86.035330} {\bibfield  {journal} {\bibinfo  {journal} {Phys.
  Rev. B}\ }\textbf {\bibinfo {volume} {86}},\ \bibinfo {pages} {035330}
  (\bibinfo {year} {2012})}\BibitemShut {NoStop}%
\bibitem [{\citenamefont {Croy}\ and\ \citenamefont
  {Saalmann}(2016)}]{Croy2016}%
  \BibitemOpen
  \bibfield  {author} {\bibinfo {author} {\bibfnamefont {A.}~\bibnamefont
  {Croy}}\ and\ \bibinfo {author} {\bibfnamefont {U.}~\bibnamefont
  {Saalmann}},\ }\bibfield  {title} {\enquote {\bibinfo {title} {Full counting
  statistics of a nonadiabatic electron pump},}\ }\href {\doibase
  10.1103/PhysRevB.93.165428} {\bibfield  {journal} {\bibinfo  {journal} {Phys.
  Rev. B}\ }\textbf {\bibinfo {volume} {93}},\ \bibinfo {pages} {165428}
  (\bibinfo {year} {2016})}\BibitemShut {NoStop}%
\bibitem [{\citenamefont {Sinitsyn}\ and\ \citenamefont
  {Nemenman}(2007{\natexlab{a}})}]{Sinitsyn2007Berry}%
  \BibitemOpen
  \bibfield  {author} {\bibinfo {author} {\bibfnamefont {N.~A.}\ \bibnamefont
  {Sinitsyn}}\ and\ \bibinfo {author} {\bibfnamefont {I.}~\bibnamefont
  {Nemenman}},\ }\bibfield  {title} {\enquote {\bibinfo {title} {The {B}erry
  phase and the pump flux in stochastic chemical kinetics},}\ }\href
  {http://stacks.iop.org/0295-5075/77/i=5/a=58001} {\bibfield  {journal}
  {\bibinfo  {journal} {Europhys. Lett.}\ }\textbf {\bibinfo {volume} {77}},\
  \bibinfo {pages} {58001} (\bibinfo {year} {2007}{\natexlab{a}})}\BibitemShut
  {NoStop}%
\bibitem [{\citenamefont {Sinitsyn}\ and\ \citenamefont
  {Nemenman}(2007{\natexlab{b}})}]{Sinitsyn2007}%
  \BibitemOpen
  \bibfield  {author} {\bibinfo {author} {\bibfnamefont {N.~A.}\ \bibnamefont
  {Sinitsyn}}\ and\ \bibinfo {author} {\bibfnamefont {I.}~\bibnamefont
  {Nemenman}},\ }\bibfield  {title} {\enquote {\bibinfo {title} {{U}niversal
  {G}eometric {T}heory of {M}esoscopic {S}tochastic {P}umps and {R}eversible
  {R}atchets},}\ }\href {\doibase 10.1103/PhysRevLett.99.220408} {\bibfield
  {journal} {\bibinfo  {journal} {Phys. Rev. Lett.}\ }\textbf {\bibinfo
  {volume} {99}},\ \bibinfo {pages} {220408} (\bibinfo {year}
  {2007}{\natexlab{b}})}\BibitemShut {NoStop}%
\bibitem [{\citenamefont {Ren}\ \emph {et~al.}(2010)\citenamefont {Ren},
  \citenamefont {H\"anggi},\ and\ \citenamefont {Li}}]{Hanggi2010}%
  \BibitemOpen
  \bibfield  {author} {\bibinfo {author} {\bibfnamefont {J.}~\bibnamefont
  {Ren}}, \bibinfo {author} {\bibfnamefont {P.}~\bibnamefont {H\"anggi}}, \
  and\ \bibinfo {author} {\bibfnamefont {B.}~\bibnamefont {Li}},\ }\bibfield
  {title} {\enquote {\bibinfo {title} {{B}erry-{P}hase-{I}nduced {H}eat
  {P}umping and {I}ts {I}mpact on the {F}luctuation {T}heorem},}\ }\href
  {\doibase 10.1103/PhysRevLett.104.170601} {\bibfield  {journal} {\bibinfo
  {journal} {Phys. Rev. Lett.}\ }\textbf {\bibinfo {volume} {104}},\ \bibinfo
  {pages} {170601} (\bibinfo {year} {2010})}\BibitemShut {NoStop}%
\bibitem [{\citenamefont {Goswami}\ \emph {et~al.}(2016)\citenamefont
  {Goswami}, \citenamefont {Agarwalla},\ and\ \citenamefont
  {Harbola}}]{Goswami2016}%
  \BibitemOpen
  \bibfield  {author} {\bibinfo {author} {\bibfnamefont {H.~P.}\ \bibnamefont
  {Goswami}}, \bibinfo {author} {\bibfnamefont {B.~K.}\ \bibnamefont
  {Agarwalla}}, \ and\ \bibinfo {author} {\bibfnamefont {U.}~\bibnamefont
  {Harbola}},\ }\bibfield  {title} {\enquote {\bibinfo {title} {Geometric
  effects in nonequilibrium electron transfer statistics in adiabatically
  driven quantum junctions},}\ }\href {\doibase 10.1103/PhysRevB.93.195441}
  {\bibfield  {journal} {\bibinfo  {journal} {Phys. Rev. B}\ }\textbf {\bibinfo
  {volume} {93}},\ \bibinfo {pages} {195441} (\bibinfo {year}
  {2016})}\BibitemShut {NoStop}%
\bibitem [{\citenamefont {Touchette}(2009)}]{TOUCHETTE2009}%
  \BibitemOpen
  \bibfield  {author} {\bibinfo {author} {\bibfnamefont {H.}~\bibnamefont
  {Touchette}},\ }\bibfield  {title} {\enquote {\bibinfo {title} {The large
  deviation approach to statistical mechanics},}\ }\href {\doibase
  https://doi.org/10.1016/j.physrep.2009.05.002} {\bibfield  {journal}
  {\bibinfo  {journal} {Phys. Rep.}\ }\textbf {\bibinfo {volume} {478}},\
  \bibinfo {pages} {1} (\bibinfo {year} {2009})}\BibitemShut {NoStop}%
\bibitem [{\citenamefont {Plenio}\ and\ \citenamefont
  {Knight}(1998)}]{Plenio1998}%
  \BibitemOpen
  \bibfield  {author} {\bibinfo {author} {\bibfnamefont {M.~B.}\ \bibnamefont
  {Plenio}}\ and\ \bibinfo {author} {\bibfnamefont {P.~L.}\ \bibnamefont
  {Knight}},\ }\bibfield  {title} {\enquote {\bibinfo {title} {The quantum-jump
  approach to dissipative dynamics in quantum optics},}\ }\href {\doibase
  10.1103/RevModPhys.70.101} {\bibfield  {journal} {\bibinfo  {journal} {Rev.
  Mod. Phys.}\ }\textbf {\bibinfo {volume} {70}},\ \bibinfo {pages} {101}
  (\bibinfo {year} {1998})}\BibitemShut {NoStop}%
\bibitem [{\citenamefont {Benito}\ \emph {et~al.}(2016)\citenamefont {Benito},
  \citenamefont {Niklas},\ and\ \citenamefont {Kohler}}]{Benito2016}%
  \BibitemOpen
  \bibfield  {author} {\bibinfo {author} {\bibfnamefont {M.}~\bibnamefont
  {Benito}}, \bibinfo {author} {\bibfnamefont {M.}~\bibnamefont {Niklas}}, \
  and\ \bibinfo {author} {\bibfnamefont {S.}~\bibnamefont {Kohler}},\
  }\bibfield  {title} {\enquote {\bibinfo {title} {Full-counting statistics of
  time-dependent conductors},}\ }\href {\doibase 10.1103/PhysRevB.94.195433}
  {\bibfield  {journal} {\bibinfo  {journal} {Phys. Rev. B}\ }\textbf {\bibinfo
  {volume} {94}},\ \bibinfo {pages} {195433} (\bibinfo {year}
  {2016})}\BibitemShut {NoStop}%
\bibitem [{\citenamefont {Pistolesi}(2004)}]{Pistolesi2004}%
  \BibitemOpen
  \bibfield  {author} {\bibinfo {author} {\bibfnamefont {F.}~\bibnamefont
  {Pistolesi}},\ }\bibfield  {title} {\enquote {\bibinfo {title} {Full counting
  statistics of a charge shuttle},}\ }\href {\doibase
  10.1103/PhysRevB.69.245409} {\bibfield  {journal} {\bibinfo  {journal} {Phys.
  Rev. B}\ }\textbf {\bibinfo {volume} {69}},\ \bibinfo {pages} {245409}
  (\bibinfo {year} {2004})}\BibitemShut {NoStop}%
\bibitem [{\citenamefont {Potanina}\ and\ \citenamefont
  {Flindt}(2017)}]{Potanina2017}%
  \BibitemOpen
  \bibfield  {author} {\bibinfo {author} {\bibfnamefont {E.}~\bibnamefont
  {Potanina}}\ and\ \bibinfo {author} {\bibfnamefont {C.}~\bibnamefont
  {Flindt}},\ }\bibfield  {title} {\enquote {\bibinfo {title} {Electron waiting
  times of a periodically driven single-electron turnstile},}\ }\href {\doibase
  10.1103/PhysRevB.96.045420} {\bibfield  {journal} {\bibinfo  {journal} {Phys.
  Rev. B}\ }\textbf {\bibinfo {volume} {96}},\ \bibinfo {pages} {045420}
  (\bibinfo {year} {2017})}\BibitemShut {NoStop}%
\bibitem [{\citenamefont {Bukov}\ \emph {et~al.}(2015)\citenamefont {Bukov},
  \citenamefont {D'Alessio},\ and\ \citenamefont
  {Polkovnikov}}]{BukovPolkovnikov2015}%
  \BibitemOpen
  \bibfield  {author} {\bibinfo {author} {\bibfnamefont {M.}~\bibnamefont
  {Bukov}}, \bibinfo {author} {\bibfnamefont {L.}~\bibnamefont {D'Alessio}}, \
  and\ \bibinfo {author} {\bibfnamefont {A.}~\bibnamefont {Polkovnikov}},\
  }\bibfield  {title} {\enquote {\bibinfo {title} {Universal high-frequency
  behavior of periodically driven systems: from dynamical stabilization to
  {F}loquet engineering},}\ }\href {\doibase 10.1080/00018732.2015.1055918}
  {\bibfield  {journal} {\bibinfo  {journal} {Adv. Phys.}\ }\textbf {\bibinfo
  {volume} {64}},\ \bibinfo {pages} {139} (\bibinfo {year} {2015})}\BibitemShut
  {NoStop}%
\bibitem [{Note1()}]{Note1}%
  \BibitemOpen
  \bibinfo {note} {We note that the left and right eigenvectors, $\protect
  \langle p_i(\chi ,t)\protect \vert $ and $\protect \vert p_i(\chi ,t)\protect
  \rangle $, are not related by simple Hermitian conjugation, since the rate
  matrix $\protect \mathbf {L}(\chi ,t)$ is not Hermitian.}\BibitemShut {Stop}%
\bibitem [{\citenamefont {Cavina}\ \emph {et~al.}(2017)\citenamefont {Cavina},
  \citenamefont {Mari},\ and\ \citenamefont {Giovannetti}}]{Cavina2017}%
  \BibitemOpen
  \bibfield  {author} {\bibinfo {author} {\bibfnamefont {V.}~\bibnamefont
  {Cavina}}, \bibinfo {author} {\bibfnamefont {A.}~\bibnamefont {Mari}}, \ and\
  \bibinfo {author} {\bibfnamefont {V.}~\bibnamefont {Giovannetti}},\
  }\bibfield  {title} {\enquote {\bibinfo {title} {Slow {D}ynamics and
  {T}hermodynamics of {O}pen {Q}uantum {S}ystems},}\ }\href {\doibase
  10.1103/PhysRevLett.119.050601} {\bibfield  {journal} {\bibinfo  {journal}
  {Phys. Rev. Lett.}\ }\textbf {\bibinfo {volume} {119}},\ \bibinfo {pages}
  {050601} (\bibinfo {year} {2017})}\BibitemShut {NoStop}%
\bibitem [{\citenamefont {Blanter}\ and\ \citenamefont
  {B\"uttiker}(2000)}]{Blanter2000}%
  \BibitemOpen
  \bibfield  {author} {\bibinfo {author} {\bibfnamefont {Ya.~M.}\ \bibnamefont
  {Blanter}}\ and\ \bibinfo {author} {\bibfnamefont {M.}~\bibnamefont
  {B\"uttiker}},\ }\bibfield  {title} {\enquote {\bibinfo {title} {Shot noise
  in mesoscopic conductors},}\ }\href {\doibase
  http://dx.doi.org/10.1016/S0370-1573(99)00123-4} {\bibfield  {journal}
  {\bibinfo  {journal} {Phys. Rep.}\ }\textbf {\bibinfo {volume} {336}},\
  \bibinfo {pages} {166} (\bibinfo {year} {2000})}\BibitemShut {NoStop}%
\bibitem [{\citenamefont {Flindt}\ \emph {et~al.}(2010)\citenamefont {Flindt},
  \citenamefont {Novotn\'y}, \citenamefont {Braggio},\ and\ \citenamefont
  {Jauho}}]{Flindt2010}%
  \BibitemOpen
  \bibfield  {author} {\bibinfo {author} {\bibfnamefont {C.}~\bibnamefont
  {Flindt}}, \bibinfo {author} {\bibfnamefont {T.}~\bibnamefont {Novotn\'y}},
  \bibinfo {author} {\bibfnamefont {A.}~\bibnamefont {Braggio}}, \ and\
  \bibinfo {author} {\bibfnamefont {A.-P.}\ \bibnamefont {Jauho}},\ }\bibfield
  {title} {\enquote {\bibinfo {title} {Counting statistics of transport through
  {C}oulomb blockade nanostructures: High-order cumulants and non-{M}arkovian
  effects},}\ }\href {\doibase 10.1103/PhysRevB.82.155407} {\bibfield
  {journal} {\bibinfo  {journal} {Phys. Rev. B}\ }\textbf {\bibinfo {volume}
  {82}},\ \bibinfo {pages} {155407} (\bibinfo {year} {2010})}\BibitemShut
  {NoStop}%
\bibitem [{\citenamefont {Blanes}\ \emph {et~al.}(2009)\citenamefont {Blanes},
  \citenamefont {Casas}, \citenamefont {Oteo},\ and\ \citenamefont
  {Ros}}]{Blanes2009}%
  \BibitemOpen
  \bibfield  {author} {\bibinfo {author} {\bibfnamefont {S.}~\bibnamefont
  {Blanes}}, \bibinfo {author} {\bibfnamefont {F.}~\bibnamefont {Casas}},
  \bibinfo {author} {\bibfnamefont {J.~A.}\ \bibnamefont {Oteo}}, \ and\
  \bibinfo {author} {\bibfnamefont {J.}~\bibnamefont {Ros}},\ }\bibfield
  {title} {\enquote {\bibinfo {title} {The {M}agnus expansion and some of its
  applications},}\ }\href {\doibase
  https://doi.org/10.1016/j.physrep.2008.11.001} {\bibfield  {journal}
  {\bibinfo  {journal} {Phys. Rep.}\ }\textbf {\bibinfo {volume} {470}},\
  \bibinfo {pages} {151} (\bibinfo {year} {2009})}\BibitemShut {NoStop}%
\bibitem [{\citenamefont {Kuwahara}\ \emph {et~al.}(2016)\citenamefont
  {Kuwahara}, \citenamefont {Mori},\ and\ \citenamefont
  {Saito}}]{KUWAHARA201696}%
  \BibitemOpen
  \bibfield  {author} {\bibinfo {author} {\bibfnamefont {T.}~\bibnamefont
  {Kuwahara}}, \bibinfo {author} {\bibfnamefont {T.}~\bibnamefont {Mori}}, \
  and\ \bibinfo {author} {\bibfnamefont {K.}~\bibnamefont {Saito}},\ }\bibfield
   {title} {\enquote {\bibinfo {title} {Floquet-{M}agnus theory and generic
  transient dynamics in periodically driven many-body quantum systems},}\
  }\href {\doibase https://doi.org/10.1016/j.aop.2016.01.012} {\bibfield
  {journal} {\bibinfo  {journal} {Ann. Phys.}\ }\textbf {\bibinfo {volume}
  {367}},\ \bibinfo {pages} {96} (\bibinfo {year} {2016})}\BibitemShut
  {NoStop}%
\bibitem [{\citenamefont {Pekola}\ \emph {et~al.}(1999)\citenamefont {Pekola},
  \citenamefont {Toppari}, \citenamefont {Aunola}, \citenamefont {Savolainen},\
  and\ \citenamefont {Averin}}]{Pekola1999}%
  \BibitemOpen
  \bibfield  {author} {\bibinfo {author} {\bibfnamefont {J.~P.}\ \bibnamefont
  {Pekola}}, \bibinfo {author} {\bibfnamefont {J.~J.}\ \bibnamefont {Toppari}},
  \bibinfo {author} {\bibfnamefont {M.}~\bibnamefont {Aunola}}, \bibinfo
  {author} {\bibfnamefont {M.~T.}\ \bibnamefont {Savolainen}}, \ and\ \bibinfo
  {author} {\bibfnamefont {D.~V.}\ \bibnamefont {Averin}},\ }\bibfield  {title}
  {\enquote {\bibinfo {title} {Adiabatic transport of {C}ooper pairs in arrays
  of {J}osephson junctions},}\ }\href {\doibase 10.1103/PhysRevB.60.R9931}
  {\bibfield  {journal} {\bibinfo  {journal} {Phys. Rev. B}\ }\textbf {\bibinfo
  {volume} {60}},\ \bibinfo {pages} {9931(R)} (\bibinfo {year}
  {1999})}\BibitemShut {NoStop}%
\bibitem [{\citenamefont {Levinson}\ \emph {et~al.}(2001)\citenamefont
  {Levinson}, \citenamefont {Entin-Wohlman},\ and\ \citenamefont
  {W{\"o}lfle}}]{Levinson2001}%
  \BibitemOpen
  \bibfield  {author} {\bibinfo {author} {\bibfnamefont {Y.}~\bibnamefont
  {Levinson}}, \bibinfo {author} {\bibfnamefont {O.}~\bibnamefont
  {Entin-Wohlman}}, \ and\ \bibinfo {author} {\bibfnamefont {P.}~\bibnamefont
  {W{\"o}lfle}},\ }\bibfield  {title} {\enquote {\bibinfo {title} {Pumping at
  resonant transmission and transferred charge quantization},}\ }\href
  {http://www.sciencedirect.com/science/article/pii/S0378437101004514}
  {\bibfield  {journal} {\bibinfo  {journal} {Physica A}\ }\textbf {\bibinfo
  {volume} {302}},\ \bibinfo {pages} {335} (\bibinfo {year}
  {2001})}\BibitemShut {NoStop}%
\end{thebibliography}
\end{document}